\documentclass[journal]{IEEEtran}

\usepackage{cite}
\usepackage{amsmath,amssymb,amsfonts}
\usepackage{graphicx}
\usepackage{textcomp}
\usepackage{xcolor}
\usepackage{graphicx}
\usepackage{comment}
\usepackage{amsmath,amsfonts}
\usepackage{bigints}
\usepackage{accents}
\usepackage{algorithm}
\usepackage{array}
\usepackage{textcomp}
\usepackage{stfloats}
\usepackage{url}
\usepackage{verbatim}
\usepackage{graphicx}
\usepackage{cite}
\usepackage{amsfonts}
\usepackage{amssymb}
\usepackage{pgfplots}
\usepackage{tikz}
\usepackage{mathtools}
\usepackage{circuitikz}
\usepackage{amsmath}
\usepackage{amsfonts}
\usepackage{amssymb}
\usepackage[amssymb]{SIunits}
\usepackage{longtable}%flexible table definitions
\usepackage{upgreek}
\usepackage{import}%to allow import of svg drawings
\usepackage{color}%to import svg drawings in color
\usepackage{algorithm,algpseudocode}% For algorithm environment
\usepackage{amsthm}
\usepackage{cite}
\usepackage{subcaption}
\usepackage{color}
\usepackage{mathequations}
\newcommand{\sminus}{\,{-}\,}

\def\pdanse{\text{pDANSE}}
\def\bse{\text{BSE}}

\def\BibTeX{{\rm B\kern-.05em{\sc i\kern-.025em b}\kern-.08em
    T\kern-.1667em\lower.7ex\hbox{E}\kern-.125emX}}
\pgfplotsset{compat=1.18}

\ifCLASSINFOpdf
\else
\fi
\begin{document}
\title{pDANSE: Particle-based Data-driven Nonlinear State Estimation from Nonlinear Measurements \\
%DANSE: Data-driven Nonlinear State Estimation of Model-free Process from Nonlinear Measurements  % Title commented out for the time being
} 

\author{
    \IEEEauthorblockN{Anubhab Ghosh \IEEEauthorrefmark{1}, Yonina C. Eldar \IEEEauthorrefmark{2},  Saikat Chatterjee \IEEEauthorrefmark{1}} 
    
    \IEEEauthorblockA{\IEEEauthorrefmark{1} Digital Futures Centre and School of Elect. Engg. $\&$ Comp. Sc., KTH Royal Institute of Technology, Sweden} \and \\
    \IEEEauthorblockA{\IEEEauthorrefmark{2} Faculty of Mathematics and Computer Science, The Weizmann Institute of Science, Israel} \\
    \IEEEauthorblockA{\IEEEauthorrefmark{1} anubhabg@kth.se \,\, \IEEEauthorrefmark{2} yonina.eldar@weizmann.ac.il \,\, \IEEEauthorrefmark{1} sach@kth.se}

    \thanks{This work has been submitted to the IEEE for future publication. Copyright may be transferred without notice, after which this version will no longer be available.  % Copyright related material for ArXiV preprint.
    The work is supported by a research grant from Digital Futures Centre. Project title: ``DLL: Data-limited learning of complex systems''. Website: https://www.digitalfutures.kth.se/. The authors would also like to mention that most of the work was carried out when Y. C. Eldar was at The Weizmann Institute of Science, Israel. Recently, in August 2025, Y. C. Eldar is also affiliated with the Department of Electrical and Computer Engineering, Northeastern University, USA (Email: y.eldar@northeastern.edu).}
}

\maketitle

\begin{abstract}

% New abstract by Saikat
We consider the problem of designing a data-driven nonlinear state estimation (DANSE) method that uses (noisy) nonlinear measurements of a process whose underlying state transition model (STM) is unknown. Such a process is referred to as a model-free process. A recurrent neural network (RNN) provides parameters of a Gaussian prior that characterize the state of the model-free process, using all previous measurements at a given time point. In the case of DANSE, the measurement system was linear, leading to a closed-form solution for the state posterior. However, the presence of a nonlinear measurement system renders a closed-form solution infeasible. Instead, the second-order statistics of the state posterior are computed using the nonlinear measurements observed at the time point. We address the nonlinear measurements using a reparameterization trick-based particle sampling approach, and estimate the second-order statistics of the state posterior. The proposed method is referred to as particle-based DANSE (pDANSE). The RNN of pDANSE uses sequential measurements efficiently and avoids the use of computationally intensive sequential Monte-Carlo (SMC) and/or ancestral sampling. 
We describe the semi-supervised learning method for pDANSE, which transitions to unsupervised learning in the absence of labeled data. Using a stochastic Lorenz-$63$ system as a benchmark process, we experimentally demonstrate the state estimation performance for four nonlinear measurement systems. We explore cubic nonlinearity and a camera-model nonlinearity where unsupervised learning is used; then we explore half-wave rectification nonlinearity and Cartesian-to-spherical nonlinearity where semi-supervised learning is used.  {Additionally, we also show the performance of pDANSE for the stochastic Lorenz-$96$ system with a half-wave, rectified measurement system. The performance of state estimation is shown to be competitive vis-\`a-vis model-driven methods that have complete knowledge of the STM of the dynamical system.} 

\end{abstract}


% \fi
\IEEEpeerreviewmaketitle

\section{Introduction}
\label{sec:intro}
Bayesian state estimation (BSE) of a nonlinear dynamical process from noisy observations (measurements) is an active area of research interest with many applications \cite{dehganpour2019survey, shlezinger2025survey}. The relationship between the state and the measurement is typically characterized using a measurement system that can be linear or nonlinear. Given a measurement system, prime examples of model-driven BSE methods are Kalman filter (KF), extended KF (EKF), unscented KF (UKF), cubature KF (CKF), and particle filter (PF). All of these approaches use a state-transition model (STM) to characterize the temporal dynamics of the process, as a-priori knowledge for BSE. An STM is typically Markovian. 
%The methods also use knowledge of measurement systems. 
KF is theoretically optimal when both the measurement system and the STM are linear Gaussian. 
%For nonlinear STMs and/or nonlinear measurement systems, EKF, UKF and PF are used. 
PF uses sequential Monte-Carlo (SMC) methods and is, {theoretically speaking, asymptotically optimal if the number of samples (or particles) is asymptotically high \cite{gustafsson2010particle}}. 
%PF is known to be computationally intensive.

Here, we consider the design of a data-driven BSE method for a model-free process where there is no knowledge of the STM, and the measurement system is nonlinear. Data-driven methods typically use machine learning techniques, mainly neural networks, and train them on the available data. Recently, the DANSE (data-driven nonlinear state estimation) method was developed and demonstrated for BSE of several nonlinear processes where the measurement system is linear Gaussian \cite{ghosh2023dansejrnl}. Our main contribution in this article is to extend DANSE for nonlinear measurement systems. 

DANSE uses a recurrent neural network (RNN) called the gated recurrent unit (GRU) \cite{choPropertiesNeuralMachine2014}. It can also use other popular RNN architectures like long short-term memory networks (LSTM) \cite{hochreiter1997long}. The RNN provides parameters of a Gaussian prior characterizing the state of the model-free process, using all previous measurements at a discrete time point. The RNN uses sequential measurements over time efficiently, maintaining causality. DANSE uses a linear Gaussian measurement system, which allows for a closed-form Gaussian posterior of the state and prediction of the measurements. The closed-form prediction of the measurements given all the previous measurements helps in training the RNN directly by formulating an appropriate maximum-likelihood based optimization function. 

For nonlinear measurements, the analytical tractability is limited and the extension of DANSE is challenging. We address the challenge using a Monte-Carlo (MC) approach - a reparameterization trick-based particle sampling technique. The approach helps to compute the second-order statistics of the posterior of the state, like a PF. Hence, the proposed method is referred to as particle DANSE (pDANSE). Our method enables computing a lower bound on the prediction of the measurements, in turn helping to train the RNN of pDANSE. The usage of RNN in pDANSE remains the same as that of DANSE, which means to provide parameters of a Gaussian prior. Unlike PF, pDANSE does not know the STM and does not use computationally intensive SMC.

We now provide a brief review of relevant methods for tackling the BSE of a nonlinear dynamical process from noisy, nonlinear measurements. In the literature, there exists a variety of methods. They can be model-driven, data-driven, or hybrid. As mentioned earlier, model-driven methods assume knowledge of the underlying STM. Notable examples are UKF, CKF, and PF \cite{julier2004unscented, wan2000unscented, arasaratnam2009cubature, gordon1993novel, arulampalam2002tutorial}. %Typically, the above methods assume that the STM is Markovian. 
The UKF uses a collection of cleverly designed sigma points and corresponding weights to represent an approximate prior distribution of the state (using second-order moments). Using the unscented transform \cite{julier2004unscented}, the sigma points are propagated using the dynamical model and subsequently the measurement system to obtain the corresponding posterior estimates. The CKF follows a similar philosophy, but instead of the unscented transform, utilizes the spherical integration rule for designing sigma points and propagating them using the SSM \cite{arasaratnam2009cubature}, \cite[Chap. 8]{sarkka2023bayesian}. The PF is sampling-based and can work with non-Gaussian dynamical systems, unlike the UKF and the CKF. It approximates the underlying prior and posterior distributions of the state using a weighted system of particles \cite{arulampalam2002tutorial}. In the simplest version of the PF, known as the Bootstrap filter \cite{gordon1993novel}, the STM is used as the proposal distribution, from which the particles at the next time instant are drawn. Then, {using the measurement system}, scalar weights for each particle are computed recursively. Finally, to avoid particle degeneracy, resampling is also carried out by sampling particles again using the computed weights \cite{doucet2001sequential}. 
{Another model-driven method suitable for tackling high-dimensional BSE problems is the ensemble Kalman filter (EnKF) \cite{roth2017ensemble}. The EnKF also uses a set of particles, referred to as \textit{ensembles}, to represent the unknown distribution. However, unlike the PF, the EnKF does not use SMC, and instead uses sub-optimal, computationally efficient Kalman-like updates for the prediction and the update step.} 
%Mention a little bit more about particle filters.

%From a theoretical standpoint, PFs are well-known to provide asymptotically optimal numerical solutions for the BSE problem \cite{miguez2004tracking}. However, it is also computationally intensive.  

%%% NOTE: Possibly, emphasize Differentiable particle filters (DPFs) here and mention that DVAE is not designed for BSE per se, but an allied approach that can be perhaps used. 
Data-driven methods use machine learning techniques and learn from available training data. The type of learning depends on the availability of data; a supervised learning approach is feasible when pairs of true state trajectories and corresponding noisy measurement trajectories are available as labelled training data, while an unsupervised learning approach uses only noisy measurement trajectories as unlabelled training data. Finally, when a limited amount of labeled data and a large amount of unlabeled data are available, semi-supervised techniques can be viable.

Notable examples of data-driven methods that do not use any STM are DANSE \cite{ghosh2023dansejrnl} and recurrent filters \cite{ruhe2021self}. Both of them assume a linear measurement system and use an RNN to provide a prior for the underlying state. Another example is the differentiable particle filter (DPF) \cite{van2025deep, corenflos2021differentiable, chen2023dpfsurvey, jonschkowski2018differentiable}. The DPF is the extension of PF, which allows an STM to be learned from data. DPF uses SMC, and a deep neural network (DNN) is used to parameterize the STM. Similarly, DNN-based STMs are used in dynamical variational autoencoder (DVAE) methods \cite{girin2021dynamical, krishnan2015deep, krishnan2017structured, fraccaro2017disentangled}. They use computationally intensive ancestral sampling and variational approximations for the posterior. The DVAE was designed for time-series modeling and can be modified for BSE. An implementation of a specific DVAE approach, known as the deep KF (DKF) \cite{krishnan2017structured} was demonstrated for BSE using linear Gaussian measurements and compared with other relevant methods in \cite[Section III.]{ghosh2023dansejrnl}. There are also Gaussian process-based (GP-based) methods for learning the STM using the kernel trick and mainly explored in system identification literature \cite{ko2009gp, frigola2013bayesian, frigola2014variational, svensson2016computationally}. The GP parameters are learned using ancestral sampling with particle Gibbs \cite{lindsten2012ancestor,lindsten2014particle}. 

%Notable examples of data-driven methods include the class of dynamical variational autoencoders (DVAEs), which model the noisy measurement sequence using a suitably learned state dynamics \cite{girin2021dynamical, krishnan2015deep, krishnan2017structured, fraccaro2017disentangled}. They do not assume any knowledge regarding the dependencies in the STM. Instead, they directly learn the STM (and the measurement system, if required) using an approximate posterior distribution in an unsupervised manner. Another class of data-driven methods uses Gaussian processes (GPs) to learn the STM. The learning approach can be supervised or unsupervised \cite{ko2009gp, frigola2013bayesian, frigola2014variational, svensson2016computationally}, and the parameters of the Gaussian processes are learned using ancestral sampling with particle Gibbs sampling \cite{lindsten2012ancestor,lindsten2014particle}. 

Hybrid methods borrow ideas from both data-driven and model-driven approaches. A notable class of hybrid methods includes the recent KalmanNet and its variations \cite{revach2022unsupervised, revach2022kalmannet, ni2024adaptive}. KalmanNet utilizes full or partial knowledge of the STM and the measurement system, and learns the Kalman gain component using training data. The learning mechanism can be supervised \cite{revach2022kalmannet} or unsupervised \cite{revach2022unsupervised}, and can handle nonlinear measurement systems.

In this article, our contribution builds on our prior work \cite{ghosh2025pdanse}, on extending DANSE for handling nonlinear measurements. We have two major additions compared to our prior work. First, while the prior work explored semi-supervised learning as the principal learning framework for $\pdanse$, in this paper, we also explore the unsupervised learning of $\pdanse$. We show how the unsupervised learning formulation of $\pdanse$ arises naturally as a sub-case of semi-supervised learning when the labelled data is unavailable. Secondly, we extend the experimental study of our prior work \cite{ghosh2025pdanse} and investigate three additional nonlinear measurement systems comprising a cubic nonlinearity, a high-dimensional, camera model nonlinearity, and a Cartesian-to-spherical nonlinearity. These are explored in addition to the half-wave rectification nonlinearity shown in the prior work \cite{ghosh2025pdanse}. The experiments are carried out using a stochastic Lorenz-$63$ system as a benchmark nonlinear dynamical process \cite{lorenz1963deterministic}. We demonstrate cubic nonlinearity and a camera-model nonlinearity where unsupervised learning is used; then we demonstrate half-wave rectification nonlinearity and Cartesian-to-spherical nonlinearity where semi-supervised learning is used. The performance of $\pdanse$ is against the model-driven PF, which knows the underlying STM exactly at the time of inference.

%Firstly, we describe the learning and inference framework of $\pdanse$ once again in this work. While the description of $\pdanse$ has considerable similarities with our prior work \cite{ghosh2025pdanse}, we importantly show that $\pdanse$ can be trained either semi-supervised / unsupervised, depending on the type of measurement system. The experiments are carried out using the Lorenz-$63$ chaotic attractor \cite{lorenz1963deterministic} as the underlying SSM for our experimental study. However, in this work, we extend the empirical study shown in \cite{ghosh2025pdanse} to show the BSE performance of $\pdanse$ on four different nonlinear measurement functions -- a cubic nonlinearity, a high-dimensional cameral model nonlinearity, a half-wave rectification nonlinearity, and finally a Cartesian-to-spherical nonlinearity. The performance of $\pdanse$ is benchmarked against the model-driven UKF and PF, which know the underlying dynamics exactly at the time of inference. 

 The outline of the paper is as follows: In Section \ref{sec:proposed_idea_pdanse}, we begin with a mathematical description of the problem formulation and provide a brief background on DANSE. We then introduce our proposed approach $\pdanse$, explaining the associated inference and learning problems. Next, in Section \ref{sec:experiments_and_results}, we describe our experimental setup, where we illustrate the BSE performance of $\pdanse$ for the stochastic Lorenz-$63$ system using the four different nonlinear measurement systems. {We also show the performance of pDANSE for the stochastic Lorenz-$96$ system with a half-wave, rectified measurement system.} Finally, in Section \ref{sec:conclusion}, we provide concluding remarks.

For consistency, we use similar notations like the {works} \cite{ghosh2023dansejrnl, ghosh2025pdanse}. %Bold font lowercase symbols denote vectors and regular lowercase font denote scalars, for example, 
Lower case $\bstate$ represents a vector while $\state_{j}$ represents the $j$'th component of $\bstate$. %A sequence of vectors $\bstate_1, \bstate_2, \ldots, \bstate_t$ is compactly denoted by $\bstate_{1:t}$. 
%Then $\state_{t,j}$ denotes the $j$'th component of $\bstate_t$. 
Upper case, bold symbols like $\bmeasmat$, represent matrices. A $T$-length sequence of vectors $\bstate_{1}, \bstate_{2}, \ldots, \bstate_{T}$ is compactly denoted as $\bstate_{1:T} \triangleq \bstate_{1}, \bstate_{2}, \ldots, \bstate_{t}, \ldots, \bstate_{T}$ and $\bstate_{1:T}^{(i)}$ denotes the $i$'th sequence; $\bstate_{t}^{(i)}$ is the $t$'th vector of the sequence $\bstate_{1:T}^{(i)}$. With a slight abuse of notation, we use $\bstate_{t}^{(l)}$ as the $l$'th sample drawn from an appropriate distribution. The operator $\left(\cdot\right)^{\top}$ denotes the transpose, $\normaldist{\cdot}{\stateMean}{\stateCov}$ represents the probability density function (pdf) of a Gaussian distribution with mean $\stateMean$ and covariance matrix $\stateCov$ and $\Ewrt{p}{\cdot}$ and $\text{tr}\left(\cdot\right)$ denotes the expectation operator w.r.t. the pdf $p$ and the trace operator respectively. The natural logarithm (with respect to base $e$) is referred to as $\log$. %The notation $\Vert \bstate \Vert^2_{\mathbf{C}}$ denotes the squared $\ell'$ norm of $\bstate$ weighted by the matrix $\mathbf{C}$, i.e. $\Vert \bstate \Vert^2_{\mathbf{C}} = \bstate^{\top} \mathbf{C} \bstate$. $\text{diag}(\mathbf{x})$ denotes a square, diagonal matrix with $\mathbf{x}$ in its main diagonal. 
$\cardinality{\Dataset}$ denotes the cardinality of the set $\Dataset$. A set of $N$ natural indices is denoted by $\IndexN{N} \triangleq \{1,2,\hdots,N \}$. The expression $\matrixexp{(\mathbf{A})}$ is the matrix exponential of $\mathbf{A}$. 
%\begin{array}{rl}
%\mathcal{N}(\mathbf{x};\mathbf{m,L}) & = \frac{1}{(2\pi)^{m/2} \, \mathrm{det}(\mathbf{L})^{1/2}} \, \mathrm{e}^{\left( - \frac{1}{2} (\mathbf{x-m})^{\top} \mathbf{L}^{-1} (\mathbf{x-m}) \right)}, \\
%\log \mathcal{N}(\mathbf{x};\mathbf{m,L}) & = -m/2 \log (2\pi) -1/2 \log \mathrm{det}(\mathbf{L})  \\
%&- \frac{1}{2} (\mathbf{x-m})^{\top} \mathbf{L}^{-1} (\mathbf{x-m}) \\
%& = -m/2 \log (2\pi) -1/2 \log \mathrm{det}(\mathbf{L})  \\
%&- \frac{1}{2} \| \mathbf{x-m} \|_{\mathbf{L}^{-1}}^2
%\end{array}
%\end{IEEEeqnarray}

%The rest of the paper is organized as follows: in section \ref{sec:proposed_idea}, we describe our proposed method, including the inference and learning problems. In section \ref{sec:experiments_and_results}, we provide a quantitative comparison of our proposed $\pdanse$ method vis-\`a-vis a model-driven PF. Finally, in section \ref{sec:conclusion} we provide some final comments and scope of future work.

\section{Particle-based DANSE ($\pdanse$)} \label{sec:proposed_idea_pdanse}
In this section, we explain the BSE problem once again mathematically and provide a brief background on DANSE, which caters to the linear measurement system. Then, we explain the inference and learning problem for the proposed method $\pdanse$. Large parts of the text in Section \ref{sec:inferenceproblem} and Section \ref{sec:learningproblem} have similarities to those in \cite[section II]{ghosh2024data}, \cite{ghosh2025pdanse}. We restate most of the derivations in this paper for completeness.

\subsection{Problem formulation}
\label{sec:problemformulation}

Let $\bstate_{t} \in \setR^{\statedim}$ be the state of a dynamical process, where $t \in \mathbb{Z}_{\geq 0}$ denotes a discrete time index.  The state is not directly observed. Instead, the state is observed through measurements $\bmeas_{t} \in \setR^{\measdim}$, as follows
\begin{equation}
\label{eq:measurementsys}
\bmeas_{t} = \bhn(\bstate_{t}) + \bmnoise_{t},
\end{equation}
for $t= 1, \ldots, T$. In \eqref{eq:measurementsys}, $\bhn: \setR^{\statedim} \to \setR^{\measdim}$ models the measurement relationship and $\bmnoise_{t} \in \setR^{\measdim}$ denotes the measurement noise with known distribution $p\left(\bmnoise_{t}\right) = \mathcal{N}\left(\bmnoise_{t}; \boldsymbol{0}, \bmnoiseCov \right)$. The measurement system can be alternatively represented by the likelihood distribution $p\left(\bmeas_{t} \vert \bstate_{t} \right) = \mathcal{N}\left(\bmeas_{t}; \bhn(\bstate_{t}), \bmnoiseCov \right)$. Throughout this article, we assume that the measurement system \eqref{eq:measurementsys} is known, i.e., we know the function $\bhn$ and the Gaussian noise covariance $\bmnoiseCov$. The Bayesian state estimation (BSE) task is to find an estimate of $\bstate_{t}$ using the sequence of past and present measurements $\bmeas_{1:t} \triangleq \bmeas_{1}, \bmeas_{2}, \hdots \bmeas_{t}$. That means, estimating the posterior $p(\bstate_{t}|\bmeas_{1:t}), \forall t$, or estimating statistical moments of $p(\bstate_{t}|\bmeas_{1:t})$ when computation of $p(\bstate_{t}|\bmeas_{1:t})$ is infeasible. This is a filtering problem. Note that we maintain causality, and the BSE can be used for many applications, for example, tracking of the dynamical process from noisy measurements.

For BSE, model-driven methods typically use a state-transition model (STM) of the dynamical process. The STM is used as a-priori knowledge. An STM characterizes the temporal dynamics of the process. A widely used STM is Markovian with additive process noise \cite{frigola2013bayesian, svensson2016computationally}
\begin{equation}\label{eq:MarkovianNLDynamicalModel}
\bstate_{t} = \bfn(\bstate_{t-1}) + \bpnoise_{t}.
\end{equation}
Here $\bfn: \setR^{\statedim} \to \setR^{\statedim}$ denotes the relationship between the current state $\bstate_{t}$ and the previous state $\bstate_{t-1}$, and $\bpnoise_{t} \in \setR^{\statedim}$ is the process noise at time $t$. In relevant literature, the measurement system \eqref{eq:measurementsys} and the STM \eqref{eq:MarkovianNLDynamicalModel} are together called the state space model (SSM). 

In this article, the dynamical process is assumed to be model-free. We do not know the STM of the process, i.e., neither the function $\bfn$ nor the distribution of $\bpnoise_{t}$. Without an STM, our task is to address the BSE in a data-driven manner.  
Here, in this article, particle-based DANSE (pDANSE) is developed for nonlinear measurements \eqref{eq:measurementsys}. 
Before delving into details regarding $\pdanse$, we first provide a brief background on DANSE, highlighting its key features and mentioning its shortcomings regarding dealing with nonlinear measurements. 

\subsection{Background: DANSE}\label{sec:danse_background}
The existing DANSE addresses the BSE problem where the measurements are linear Gaussian. %That means the function $\bhn$ is linear. 
This means that in \eqref{eq:measurementsys}, we have $\bhn(\bstate_t) = \bmeasmat \bstate_{t}$ in the case of DANSE. %DANSE uses unsupervised learning employing a training dataset comprising linear Gaussian measurements. 
The core of the DANSE method is based on parameterizing a prior distribution $p\left(\bstate_{t} \vert \bmeas_{1:t-1} \right)$ as a Gaussian distribution with parameters from an RNN \cite{ghosh2023dansejrnl}. Owing to the causal nature of the state estimation task, the RNN operates in unidirectional mode, using measurements $\bmeas_{1:t-1}$ and providing the Gaussian prior parameters for $\bstate_{t}$. The prior distribution is then as follows
\begin{eqnarray}
\label{eq:prior_lik}
\begin{array}{c}
p(\bstate_t \vert \bmeas_{1:t-1}; \rnnParam) \!=\! \normaldist {\bstate_t}{\stateMeanprior{t}(\rnnParam)}{\stateCovprior{t}(\rnnParam) }, \\
\text{s.t. } \lbrace \stateMeanprior{t}(\rnnParam), \stateCovprior{t}(\rnnParam) \rbrace \triangleq \mathrm{RNN} (\bmeas_{1:t-1}; \rnnParam). \\
\end{array}
\end{eqnarray}
Here, $\stateMeanprior{t}(\rnnParam) \in \setR^{\statedim}$ and $\stateCovprior{t}(\rnnParam) \in \setR^{\statedim \times \statedim}$ denote the mean and covariance matrix of the Gaussian prior distribution, respectively. The covariance matrix is modelled as a diagonal for convenience. 
After characterising the prior, we can compute the posterior $p(\bstate_t|\bmeas_{1:t}; \rnnParam)$ in closed-form Gaussian distribution using the linear measurement setup. Recall that this means that in \eqref{eq:measurementsys}, we have $p\left(\bmeas_{t} \vert \bstate_{t} \right) = \normaldist{\bmeas_{t}}{\bmeasmat \bstate_{t}}{\bmnoiseCov}$ in the case of DANSE. The posterior is obtained by 
using the `completing the square' approach \cite[Chap. 2]{bishop2006pattern} as
\begin{eqaligned}\label{eq:posterior_update}
&p(\bstate_t \vert \bmeas_{1:t}; \rnnParam) \\
&= \frac{p(\bmeas_t \vert \bstate_{t})p(\bstate_t \vert \bmeas_{1:t-1}; \rnnParam)}{p(\bmeas_t \vert \bmeas_{1:t-1}; \rnnParam)} \\
&= \frac{p(\bmeas_t \vert \bstate_{t})p(\bstate_t \vert \bmeas_{1:t-1}; \rnnParam)}{\int p(\bmeas_t \vert \bstate^{'}_{t})p(\bstate^{'}_t \vert \bmeas_{1:t-1}; \rnnParam) d \bstate^{'}_{t}} \\
&= \frac{\normaldist{\bmeas_{t}}{\bmeasmat\bstate_{t}}{\bmnoiseCov}\normaldist {\bstate_t}{\stateMeanprior{t}(\rnnParam)}{\stateCovprior{t}(\rnnParam) }}{\int \normaldist{\bmeas_{t}}{\bmeasmat\bstate^{'}_{t}}{\bmnoiseCov}\normaldist {\bstate^{'}_t}{\stateMeanprior{t}(\rnnParam)}{\stateCovprior{t}(\rnnParam) } d \bstate^{'}_{t}} \\
&= \frac{\normaldist{\bmeas_{t}}{\bmeasmat\bstate_{t}}{\bmnoiseCov}\normaldist {\bstate_t}{\stateMeanprior{t}(\rnnParam)}{\stateCovprior{t}(\rnnParam) }}{\normaldist{\bmeas_{t}}{\bmeasmat \stateMeanprior{t}}{ \bmeasmat \stateCovprior{t} \bmeasmat^{\top} + \bmnoiseCov}} \\
&=  \normaldist{\bstate_t}{\stateMeanposterior{t}(\rnnParam)}{\stateCovposterior{t}(\rnnParam)}, 
\end{eqaligned}
where the posterior mean and covariance matrix at time $t$ are
\begin{eqaligned}\label{eq:posteriormeanscov}
\stateMeanposterior{t}(\rnnParam) &= \stateMeanprior{t}(\rnnParam) + \mathbf{K}_{t\vert 1:t-1}(\rnnParam) \pmb{\varepsilon}_t(\rnnParam), \\
\stateCovposterior{t}(\rnnParam) %& = ([\stateCovprior{t}(\pmb{\phi})]^{-1} + \bmeasmat^{\top}  \bmnoiseCov^{-1} \bmeasmat )^{-1} \\
&= \stateCovprior{t}(\rnnParam) -  \mathbf{K}_{t\vert 1:t-1}(\rnnParam)\mathbf{R}_{\varepsilon, t}(\rnnParam) \mathbf{K}_{t\vert 1:t-1}^{\top}(\rnnParam),\\
%\end{aligned}
%\end{equation}
\end{eqaligned}
with
\begin{equation}\label{eq:posterior_update_additional}
    \begin{array}{l}
    \mathbf{K}_{t\vert 1:t-1}(\rnnParam)  \triangleq \stateCovprior{t}(\rnnParam)\bmeasmat^{\top}\mathbf{R}_{\varepsilon, t}^{-1}(\rnnParam), \\ \mathbf{R}_{\varepsilon, t}(\rnnParam)  \triangleq \bmeasmat\stateCovprior{t}\left(\rnnParam\right)\bmeasmat^{\top} + \bmnoiseCov, \,\, \mathrm{and} \\
    {\pmb{\varepsilon}}_t(\rnnParam)  \triangleq \bmeas_t - \bmeasmat\stateMeanprior{t}(\rnnParam). 
    \end{array}
\end{equation}
The parameters of DANSE $\rnnParam$ are learned by maximizing the joint (log) likelihood of the measurements $p\left(\bmeas_{1:T}\right)$, where by the product rule $p\left(\bmeas_{1:T}; \rnnParam\right) = \prod_{t=1}^{T}
p\left( \bmeas_{t} | \bmeas_{1:t-1} ; \rnnParam\right)$ \cite{ghosh2023dansejrnl}. The conditional distribution $p\left( \bmeas_{t} | \bmeas_{1:t-1}; \rnnParam\right)$ can be found in closed form as
\begin{eqaligned}
\label{eq:pyt_given_prev}
&p(\bmeas_t | \bmeas_{1:t-1}; \rnnParam) \\
%&= \displaystyle\int p(\bmeas_t , \bstate_t  |  \bmeas_{1:t-1} ; \rnnParam) \, d\bstate_t \\
&= \displaystyle\int p(\bmeas_t | \bstate_t )p(\bstate_t |  \bmeas_{1:t-1} ; \rnnParam) \, d\bstate_t \\
&= \mathcal{N}\left(\bmeas_t ; \bmeasmat \stateMeanprior{t}(\rnnParam), \bmnoiseCov + \bmeasmat \stateCovprior{t}(\rnnParam) \bmeasmat^{\top}\right). \\
 %\triangleq p(\bmeas_t | \bmeas_{1:t-1}; \rnnParam ).
\end{eqaligned}
In the above, we used \eqref{eq:prior_lik}, and the sum-rule for finding Gaussian marginals as in \cite[Chap. 2]{bishop2006pattern}. 
Now, we can formulate a maximum-likelihood based optimization problem for learning the parameters $\rnnParam$ using noisy measurements, as shown in \cite[Section II-C]{ghosh2023dansejrnl}. The overall optimization problem 
is non-convex and is solved using gradient descent. 

Despite these encouraging features, we note that one of the key aspects of DANSE -- the closed form of the posterior distribution $p(\bstate_t \vert \bmeas_{1:t}; \rnnParam)$ fails to hold in the case of a nonlinear measurement system like \eqref{eq:measurementsys}, since $p(\bstate_t \vert \bmeas_{1:t}; \rnnParam)$ becomes non-Gaussian. Neither does the marginal distribution $p(\bmeas_t | \bmeas_{1:t-1}; \rnnParam)$ remain Gaussian either, prohibiting a direct optimization of the RNN parameters $\rnnParam$. In the next subsection, we explain how we address these shortcomings of DANSE. 

\subsection{Inference problem (Filtering): $\pdanse$}
\label{sec:inferenceproblem}
Let us now focus once again on the BSE problem with nonlinear measurements described in Section \ref{sec:problemformulation}. The inference problem (filtering) comprises finding the posterior distribution $p\left(\bstate_t \vert \bmeas_{1:t}\right)$ at time $t$. Note that we assume that we do not know the underlying STM. Like DANSE \cite{ghosh2023dansejrnl}, we assume that the prior distribution of $\bstate_{t}$ given $\bmeas_{1:t-1}$ in $\pdanse$ is also Gaussian (like in \eqref{eq:prior_lik}). The parameters of the Gaussian prior, that is, the mean and covariance, are given by an RNN with learnable parameters $\rnnParam$ as follows -
\begin{IEEEeqnarray}{rl}
\label{eq:RNNPrior}
& \mathrm{Prior}:  
     p\left(\bstate_t \vert \bmeas_{1:t-1}; \rnnParam\right) = \mathcal{N}\!\left(\bstate_t; \stateMeanprior{t}\left(\rnnParam\right), \stateCovprior{t}\left(\rnnParam\right)\right), \nonumber \\
    &\left\lbrace \stateMeanprior{t}\left(\rnnParam\right), \stateCovprior{t}\left(\rnnParam\right) \right\rbrace = \mathrm{RNN}\left(\bmeas_{1:t-1};\rnnParam\right),
\end{IEEEeqnarray}
where $\stateMeanprior{t}(\rnnParam) \in \setR^{\statedim}$ and $\stateCovprior{t}(\rnnParam) \in \setR^{\statedim \times \statedim}$ are obtained using an RNN having the structure shown in \cite[section II-A]{ghosh2024data}. We use a diagonal covariance matrix $\stateCovprior{t}(\rnnParam)$. Then, using the measurement system \eqref{eq:measurementsys}, we seek to compute the posterior $p\left(\bstate_t \vert \bmeas_{1:t}; \rnnParam \right)$. However, since \eqref{eq:measurementsys} is nonlinear, the likelihood distribution is $p\left(\bmeas_{t} \vert \bstate_{t} \right) = \mathcal{N}\left(\bmeas_{t}; \bhn(\bstate_{t}), \bmnoiseCov \right)$, resulting in $p\left(\bstate_t \vert \bmeas_{1:t}; \rnnParam \right)$ being intractable. One can note this by observing that 
\begin{IEEEeqnarray}{rl}
\label{eq:intractable_posterior}
&p\left(\bstate_t \vert \bmeas_{1:t}; \rnnParam \right) \nonumber\\
&= \frac{p\left(\bmeas_t \vert \bstate_{t} \right) p\left(\bstate_t \vert \bmeas_{1:t-1}; \rnnParam \right)}{p\left(\bmeas_t \vert \bmeas_{1:t-1}; \rnnParam \right)} \nonumber\\
&= \frac{p\left(\bmeas_t \vert \bstate_{t} \right) p\left(\bstate_t \vert \bmeas_{1:t-1}; \rnnParam \right)}{\int p\left(\bmeas_t \vert \bstate^{'}_{t} \right) p\left(\bstate^{'}_{t} \vert \bmeas_{1:t-1}; \rnnParam \right) d\bstate^{'}_{t}} \\
&= \frac{\mathcal{N}\left(\bmeas_{t}; \bhn(\bstate_{t}), \bmnoiseCov \right) \mathcal{N}\left(\bstate_{t}; \stateMeanprior{t}\left(\rnnParam\right), \stateCovprior{t}\left(\rnnParam\right)\right)}{\int \mathcal{N}\left(\bmeas_{t}; \bhn(\bstate^{'}_{t}), \bmnoiseCov \right) \mathcal{N}\left(\bstate^{'}_{t}; \stateMeanprior{t}\left(\rnnParam\right), \stateCovprior{t}\left(\rnnParam\right)\right) d\bstate^{'}_{t}}. \nonumber
\end{IEEEeqnarray}
From \eqref{eq:intractable_posterior}, it can be seen that the nonlinear function $\bhn(\cdot)$ renders the posterior distribution non-Gaussian. It hinders the closed-form computation of the integral in the denominator and an exact posterior distribution as in \eqref{eq:posterior_update} and \eqref{eq:posterior_update_additional}. 

\begin{figure}[t]
    \centering
    \includegraphics[width=\linewidth]{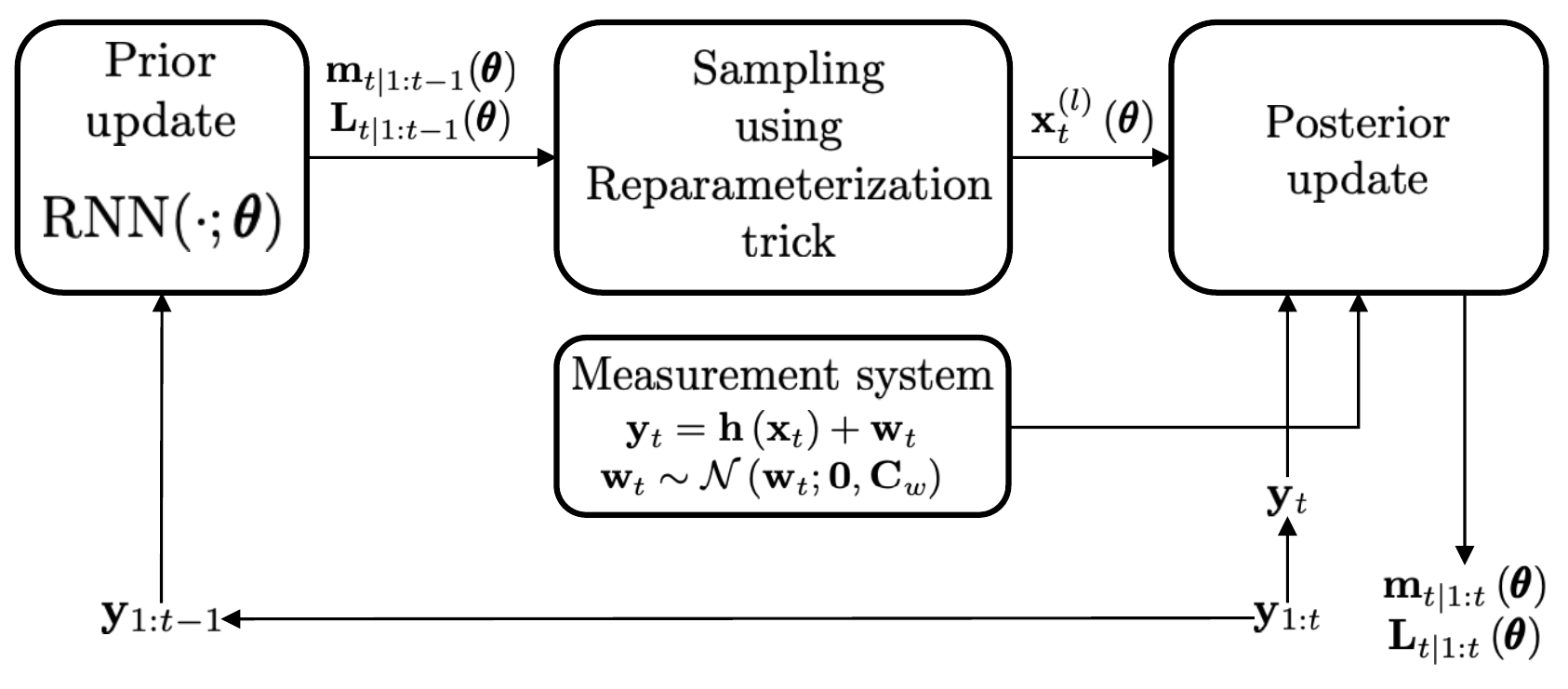}
    \caption{Schematic of $\pdanse$ at inference time.}
    \label{fig:pdanse_inference}
\end{figure}

Instead of the closed-form posterior, we compute moments of the posterior $p\left(\bstate_t \vert \bmeas_{1:t}; \rnnParam \right)$, namely the first and second-order statistical moments, using Monte-Carlo (MC) approximations. Let $\bphi: \setR^{m} \to \mathcal{M}$ denote the moment function that maps the state $\bstate_{t}$ to the space of moments $\mathcal{M}$. Note that in the case of the mean vector, $\mathcal{M} = \setR^{\statedim}$, while for the covariance matrix, $\mathcal{M} = \setR^{\statedim \times \statedim}$. The moments can be determined by solving the following expectation
\begin{IEEEeqnarray}{RL}
    &\Ewrt{p\left(\bstate_t \vert \bmeas_{1:t}; \rnnParam \right)}{\bphi\left(\bstate_{t}\right)} = \int \bphi\left(\bstate_{t}\right) p\left(\bstate_t \vert \bmeas_{1:t}; \rnnParam\right) d \bstate_{t} \nonumber\\
    &= \int \bphi\left(\bstate_{t}\right) \frac{p\left(\bmeas_t \vert \bstate_t\right)p\left(\bstate_t \vert \bmeas_{1:t-1}; \pmb{\theta}\right)}{p\left(\bmeas_t \vert \bmeas_{1:t-1} ; \rnnParam\right)} d \bstate_{t} \nonumber\\
    &= \int \bphi\left(\bstate_{t}\right) \frac{p\left(\bmeas_t \vert \bstate_t\right)p\left(\bstate_t \vert \bmeas_{1:t-1}; \pmb{\theta}\right)}{\int p\left(\bmeas_t \vert \bstate^{'}_t\right)p\left(\bstate^{'}_t \vert \bmeas_{1:t-1}; \pmb{\theta}\right) d \bstate^{'}_{t}} d \bstate_{t} \nonumber\\
    %&\approx \sum\limits_{l=1}^{L}  \bphi\left(\bstate_{t}^{(l)}(\rnnParam)\right) \frac{p\left(\bmeas_t \vert \bstate^{(l)}_t(\rnnParam)\right)}{\sum_{l' =  1}^{L} p\left(\bmeas_t \vert \bstate^{(l')}_t(\rnnParam)\right)}  \nonumber\\
    &\approx \sum\limits_{l=1}^{L}  \bphi\left(\bstate_{t}^{(l)}(\rnnParam)\right) w_{t}^{(l)}(\rnnParam), \label{eq:approxposteriormoments}
\end{IEEEeqnarray}
where we have particles sampled from the prior distribution $p\left(\bstate_t \vert \bmeas_{1:t-1}; \rnnParam\right) = \mathcal{N}\!\left(\bstate_t; \stateMeanprior{t}\left(\rnnParam\right), \stateCovprior{t}\left(\rnnParam\right)\right)$ as follows 
\begin{IEEEeqnarray}{rl}\label{eq:inferencetimesampling}
    \bstate_t^{(l)}\left(\rnnParam\right) 
    &\sim \mathcal{N}\!\left(\bstate_t; \stateMeanprior{t}\left(\rnnParam\right), \stateCovprior{t}\left(\rnnParam\right)\right), \nonumber \\
    &= \stateMeanprior{t}(\rnnParam) + \stateCovprior{t}^{{1}/{2}}(\rnnParam) \boldsymbol{\epsilon}^{(l)},
\end{IEEEeqnarray}
%$\bstate_t^{(l)} \sim p\left(\bstate_t \vert \bmeas_{1:t-1}; \rnnParam\right) = \mathcal{N}\!\left(\bstate_t; \stateMeanprior{t}\left(\rnnParam\right), \stateCovprior{t}\left(\rnnParam\right)\right)$ 
where $\boldsymbol{\epsilon}^{(l)} \sim \normaldist{\boldsymbol{\epsilon}}{\boldsymbol{0}}{\mathbf{I}}$, for $l=1, \ldots, L$. $L$ denotes the number of MC samples, and the superscript $l$ is explicitly used as the sampling index. The samples are drawn in \eqref{eq:inferencetimesampling} using the reparameterization trick \cite{kingma2013auto}. The posterior weights $w_{t}^{(l)}(\rnnParam)$ for each particle in \eqref{eq:approxposteriormoments} are 
\begin{IEEEeqnarray}{rl}
w_{t}^{(l)}(\rnnParam) &= \dfrac{p\left(\bmeas_t \vert \bstate^{(l)}_t(\rnnParam)\right)}{\sum_{l' =  1}^{L} p\left(\bmeas_t \vert \bstate^{(l')}_t(\rnnParam)\right)} \nonumber \\
%w_{t}^{(l)} &= {p\left(\bmeas_t \vert \bstate^{(l)}_t\right)} \bigg \slash {\sum_{l' =  1}^{L} p\left(\bmeas_t \vert \bstate^{(l')}_t\right)},
&= \frac{\mathcal{N}\!\left(\bmeas_t; \bhn\left(\bstate_t^{(l)}(\rnnParam)\right), \bmnoiseCov \right)}{\sum_{l' =  1}^{L} \mathcal{N}\!\left(\bmeas_t; \bhn\left(\bstate_t^{(l')}(\rnnParam)\right), \bmnoiseCov \right)},
\label{eq:weights}
\end{IEEEeqnarray}
where $\sum_{l=1}^{L} w_{t}^{(l)}(\rnnParam) = 1, \forall t$ and using the Gaussian measurement system in \eqref{eq:measurementsys}. In practice, \eqref{eq:weights} is computed in the log-domain (for numerical stability), with the denominator computed using the well-known log-sum-exp trick \cite{blanchard2021accurately} 
\begin{IEEEeqnarray}{rl}
\label{eq:weights_logsumexp}
&\log w_{t}^{(l)}(\rnnParam) \nonumber\\
&= \log \left(\dfrac{p\left(\bmeas_t \vert \bstate^{(l)}_t(\rnnParam)\right)}{\sum_{l' =  1}^{L} p\left(\bmeas_t \vert \bstate^{(l')}_t(\rnnParam)\right)}\right) \nonumber \\
%w_{t}^{(l)} &= {p\left(\bmeas_t \vert \bstate^{(l)}_t\right)} \bigg \slash {\sum_{l' =  1}^{L} p\left(\bmeas_t \vert \bstate^{(l')}_t\right)},
&= \log {\mathcal{N}\!\left(\bmeas_t; \bhn\left(\bstate_t^{(l)}(\rnnParam)\right), \bmnoiseCov \right)} \nonumber\\
&- \log {\sum_{l' =  1}^{L} \exp \log \mathcal{N}\!\left(\bmeas_t; \bhn\left(\bstate_t^{(l')}(\rnnParam)\right), \bmnoiseCov \right)}, \nonumber \\
&= \log {\mathcal{N}\!\left(\bmeas_t; \bhn\left(\bstate_t^{(l)}(\rnnParam)\right), \bmnoiseCov \right)} \nonumber\\
&- \textrm{ LSE}\left(\left\lbrace\log\mathcal{N}\!\left(\bmeas_t; \bhn\left(\bstate_t^{(l')}(\rnnParam)\right), \bmnoiseCov \right)\right\rbrace_{l'=1}^{L}\right). 
\end{IEEEeqnarray}
Here $\textrm{LSE}(\left\lbrace \varrho^{(l)} \right\rbrace_{l=1}^{L}) = \log \sum_{l=1}^{L} \exp \varrho^{(l)}$ denotes the log-sum-exp (LSE) function, for a collection of $L$ real-valued scalars $\left\lbrace \varrho^{(l)} \right\rbrace_{l=1}^{L}$. 
Using \eqref{eq:approxposteriormoments}, \eqref{eq:weights}, and \eqref{eq:weights_logsumexp}, we can obtain posterior mean and covariance estimates. Specifically, if we set $\bphi\left(\bstate_t\right) = \bstate_{t}$ in \eqref{eq:approxposteriormoments}, we obtain a posterior mean estimate $\stateMeanposterior{t}\left(\rnnParam\right)$ as follows
\begin{IEEEeqnarray}{RL}
     \stateMeanposterior{t}\left(\rnnParam\right) &= \Ewrt{p\left(\bstate_t \vert \bmeas_{1:t}; \rnnParam \right)}{\bstate_{t}}
    %&\approx \sum\limits_{l=1}^{L} \bstate_{t}^{(l)} \frac{p\left(\bmeas_t \vert \bstate^{(l)}_t\right)}{\sum_{l' =  1}^{L} p\left(\bmeas_t \vert \bstate^{(l')}_t\right)} \\
    \approx \sum_{l=1}^{L} \bstate_{t}^{(l)} (\rnnParam) w_{t}^{(l)}(\rnnParam). 
\label{eq:approxposteriormean}
\end{IEEEeqnarray}
Similarly, we get a posterior covariance estimate $\stateCovposterior{t}(\rnnParam)$ using $\bphi\left(\bstate_t\right) = \bvarepsilon_t \bvarepsilon_t^{\top}, \bvarepsilon_t \triangleq \bstate_{t} - \stateMeanposterior{t}\left(\rnnParam\right)$ in \eqref{eq:approxposteriormoments} 
\begin{IEEEeqnarray}{RL}
    \stateCovposterior{t}\left(\rnnParam\right) 
    %&= \Ewrt{p\left(\bstate_t \vert \bmeas_{1:t}; \rnnParam \right)}{\left( \bstate_{t} - \stateMeanposterior{t}\left(\rnnParam\right) \right)\left(\bstate_{t} 
    %- \stateMeanposterior{t}\left(\rnnParam\right)\right)^{\top}} \\
    &= \Ewrt{p\left(\bstate_t \vert \bmeas_{1:t}; \rnnParam \right)}{\bvarepsilon_t \bvarepsilon_t^{\top}} \nonumber \\
    %\nonumber
    %&\approx \sum\limits_{l=1}^{L} \bstate_{t}^{(l)} \frac{p\left(\bmeas_t \vert \bstate^{(l)}_t\right)}{\sum_{l' =  1}^{L} p\left(\bmeas_t \vert \bstate^{(l')}_t\right)} \\
    %&\approx \sum
    %_{l=1}^{L} \left( \bstate_{t}^{(l)} - \stateMeanposterior{t}\left(\rnnParam\right) \right)\left(\bstate_{t}^{(l)} 
 %- \stateMeanposterior{t}\left(\rnnParam\right)\right)^{\top} w_{t}^{(l)}, \\ 
    %&\approx \sum
    %_{l=1}^{L} w_{t}^{(l)} \left( \bstate_{t}^{(l)} - %\stateMeanposterior{t}\left(\rnnParam\right) \right)\left(\bstate_{t}^{(l)} 
    %- \stateMeanposterior{t}\left(\rnnParam\right)\right)^{\top}, \\ 
    &\approx \sum
    _{l=1}^{L}\bvarepsilon_t^{(l)}(\rnnParam)  \left({\bvarepsilon_t^{(l)}(\rnnParam) }\right)^{\top} w_{t}^{(l)}(\rnnParam). 
\label{eq:approxposteriorcov}
\end{IEEEeqnarray}
In \eqref{eq:approxposteriorcov}, we have $\bvarepsilon^{(l)}_t(\rnnParam) = \bstate^{(l)}_{t}(\rnnParam) - \stateMeanposterior{t}\left(\rnnParam\right)$. Also in \eqref{eq:approxposteriormean} and \eqref{eq:approxposteriorcov}, $\bstate_t^{(l)}(\rnnParam) $ are sampled from $\mathcal{N}\!\left(\bstate_t; \stateMeanprior{t}\left(\rnnParam\right), \stateCovprior{t}\left(\rnnParam\right)\right)$ for $l=1, \ldots, L$ and the corresponding weights $w_{t}^{(l)}(\rnnParam)$ are computed in log-domain according to \eqref{eq:weights_logsumexp} $\forall t$. A complete schematic of the operation of $\pdanse$ at inference time is shown in Fig. \ref{fig:pdanse_inference}.

{Finally, to clearly distinguish the inference procedure from PFs, we remark that pDANSE does not involve any form of resampling and propagation of particles from the posterior to the prior distribution - phenomena that are quintessential in the case of model-driven PFs. Instead, fresh particles are drawn from the RNN-based prior distribution using the reparameterization trick in \eqref{eq:inferencetimesampling}. This is shown in Fig. \ref{fig:pdanse_inference}.}
\subsection{Learning problem: $\pdanse$}\label{sec:learningproblem}
The learning problem concerns optimizing the parameters $\rnnParam$ of the RNN used in \pdanse. Here, we will show that $\pdanse$ can be trained in an unsupervised manner, and, if required, alternatively in a semi-supervised manner. For the ease of illustration, we begin with the description of the semi-supervised learning setup and later show how one can obtain a purely unsupervised learning setup as a sub-case in the absence of labelled training data. For a linear $\bhn$ and compressed measurements, we have previously shown the utility of semi-supervised learning in our prior work on Bayesian state estimation using compressed, linear measurements \cite{ghosh2024data}. %Here, we note that $\bhn$ is nonlinear, hinting at the fact that semi-supervised learning might be useful in this case as well. 
 %We provide the complete mathematical derivations. The derivations have similarities with those found in \cite[Section II.]{ghosh2025pdanse} with important differences. for completeness, while providing more detailed clarifications wherever possible. 
 
In a semi-supervised learning scenario, one typically has access to a limited amount of labelled training data and a significantly larger amount of unlabelled training data \cite[Section III-A]{ghosh2024data}. Let there be two disjoint sets of indices $\IndexSet_{s}$ and $\IndexSet_{u}$ for supervised and unsupervised training datasets, respectively. The training dataset required for semi-supervised learning is denoted by $\Dataset_{\text{semi}}$. $\Dataset_{\text{semi}}$ consists of two parts -- a labelled dataset $\Dataset_{s}$ having cardinality $\vert \IndexSet_{s} \vert \triangleq N_{s}$ and an unlabelled dataset $\Dataset_{u}$ having cardinality $\vert \IndexSet_{u} \vert \triangleq N_{u}$, s.t. $N_s \ll N_u \leq N = N_s + N_u$ and $\IndexSet_{s} \cup \IndexSet_{u} = \left\lbrace 1, 2, \ldots, N \right\rbrace$. Specifically, $\Dataset_{s} = \left\lbrace \left(\bstate_{1:T}^{(j)}, \bmeas_{1:T}^{(j)} \right)\right\rbrace_{j \in \IndexSet_{s}}$ consists of pairwise state and noisy measurement trajectories, having the same length $T$. Note that the sequences could have varying lengths. Here we use the same length across sequences for notational clarity and later ease of implementation. Analogously, $\Dataset_{u} = \left\lbrace \bmeas_{1:T}^{(i)}\right\rbrace_{i \in \IndexSet_{u}}$ consists of noisy measurement trajectories. Note that the superscripts $i$ and $j$ are explicitly used as sequence indices and not to be confused with the superscript $l$ used as sampling index. We can learn the parameters $\rnnParam$ by solving the following maximum-likelihood-based optimization problem: %{(Where are $N_s$, $N_u$, $N$ quantities?)} % Anubhab: Now, added in the text

\begin{IEEEeqnarray}{rl}
\rnnParam^\star &= \arg\max\limits_{\rnnParam} \log p\left({\Dataset_{\text{semi}}}; \rnnParam\right) \nonumber\\
&=\arg\max\limits_{\rnnParam} \log p\left({\Dataset_{{s}}}, {\Dataset_{{u}}}; \rnnParam\right) \nonumber\\
%&=\arg\max\limits_{\rnnParam} \log p\left(\left\lbrace \left(\bstate_{1:T}^{(j)}, \bmeas_{1:T}^{(j)}\right)\right\rbrace_{j \in \IndexSet_{s}}, \left\lbrace \bmeas_{1:T}^{(i)}\right\rbrace_{i \in \IndexSet_{u}}; \rnnParam\right) \nonumber \\
&=\arg\max\limits_{\rnnParam}  \log p\left(\left\lbrace \left(\bstate_{1:T}^{(i)} \Big\vert \bmeas_{1:T}^{(j)} \right)\right\rbrace_{j \in \IndexSet_{s}}; \rnnParam\right) \nonumber \\
&+ \log p\left(\left\lbrace \bmeas_{1:T}^{(i)}\right\rbrace_{i \in \IndexSet_{s}\cup\IndexSet_{u}}; \rnnParam\right).
\label{eq:semi-supervised-learning-problem}
\end{IEEEeqnarray}
\begin{figure}[t]
    \centering
    \includegraphics[width=\linewidth]{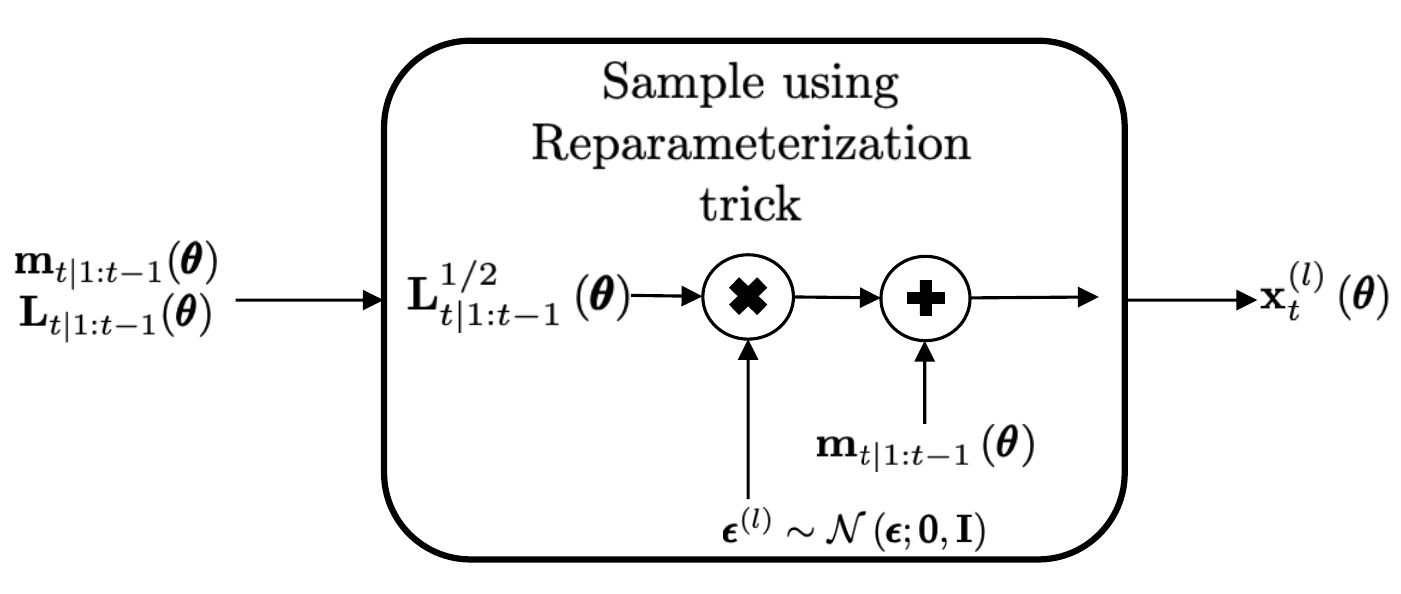}
    \caption{Sampling from the Gaussian prior in \eqref{eq:RNNPrior} using the reparameterization trick, to obtain samples $\lbrace \bstate^{(l)}_{t}\left(\rnnParam\right)\rbrace_{l=1}^{L}$.}
    \label{fig:reparam_trick}
\end{figure}
From \eqref{eq:semi-supervised-learning-problem}, the loss function consists of two terms - one constituted using only $\Dataset_{s}$ and the other using measurement trajectories from $\Dataset_{s} \cup \Dataset_{u}$. We now focus on further simplification of the above loss function in the last line of \eqref{eq:semi-supervised-learning-problem}. To start with, dropping the superscript $i$, we notice that the logarithm of the joint marginal distribution $p\left(\bmeas_{1:t}; \rnnParam\right)$ can be simplified according to the product rule, yielding 
\begin{IEEEeqnarray}{rl}
\log p\left(\bmeas_{1:T}; \rnnParam\right) &= \log \prod_{t=1}^{T} p\left(\bmeas_{t} \vert \bmeas_{1:t-1}; \rnnParam\right) \nonumber \\
&= \sum_{t=1}^{T} \log p\left(\bmeas_{t} \vert \bmeas_{1:t-1}; \rnnParam\right).
\label{eq:marginal-dist-decomposition}
\end{IEEEeqnarray}
Thus, we need to obtain the logarithm of the factorized marginal distribution $p\left(\bmeas_{t} \vert \bmeas_{1:t-1}; \rnnParam\right)$. However, the nonlinear measurement function $\bhn$ prohibits a closed-form expression of $\log p\left(\bmeas_{t} \vert \bmeas_{1:t-1}; \rnnParam\right)$. This is because
\begin{eqnarray}
\begin{array}{l}
\log p\left(\bmeas_{t} \vert \bmeas_{1:t-1}; \rnnParam\right) \\
= \log \displaystyle \int p\left(\bmeas_t \vert \bstate_{t} \right) p\left(\bstate_{t} \vert \bmeas_{1:t-1}; \rnnParam \right) d\bstate_{t} \\
= \log \Ewrt{p\left(\bstate_{t} \vert \bmeas_{1:t-1}; \rnnParam \right)}{\mathcal{N}\left(\bmeas_{t}; \bhn(\bstate_{t}), \bmnoiseCov \right)}
\end{array}
\label{eq:marginal-dist-intractable}
\end{eqnarray}
has analytically intractable expectation of the likelihood ${\mathcal{N}\left(\bmeas_{t}; \bhn(\bstate_{t}), \bmnoiseCov \right)}$ with respect to the prior distribution $p\left(\bstate_{t} \vert \bmeas_{1:t-1}; \rnnParam \right)$. Instead of directly optimizing $\log p\left(\bmeas_{t} \vert \bmeas_{1:t-1}; \rnnParam\right)$, we optimize the following lower bound 
\begin{IEEEeqnarray}{rl}
&\log p\left(\bmeas_{t} \vert \bmeas_{1:t-1}; \rnnParam\right) \nonumber \\
&=\log \Ewrt{p\left(\bstate_{t} \vert \bmeas_{1:t-1}; \rnnParam \right)}{\mathcal{N}\left(\bmeas_{t}; \bhn(\bstate_{t}), \bmnoiseCov \right)} \nonumber\\
&\geq \Ewrt{p\left(\bstate_{t} \vert \bmeas_{1:t-1}; \rnnParam \right)}{\log \mathcal{N}\left(\bmeas_{t}; \bhn(\bstate_{t}), \bmnoiseCov \right)} \nonumber\\
&\approx \frac{1}{L} \sum_{l=1}^{L} \log\mathcal{N}\left(\bmeas_{t}; \bhn\left(\bstate^{(l)}_{t}\left(\rnnParam\right)\right), \bmnoiseCov \right).
\label{eq:marginal-dist-LB}
\end{IEEEeqnarray}
The inequality in the second line of \eqref{eq:marginal-dist-LB} is obtained using Jensen's inequality and the resulting expectation is approximated in the third line by an MC approximation with samples $\bstate^{(l)}_{t}\left(\rnnParam\right)$ sampled from the Gaussian prior $p\left(\bstate_{t} \vert \bmeas_{1:t-1}; \rnnParam \right) = \mathcal{N}\left(\bstate_{t}; \stateMeanprior{t}(\rnnParam), \stateCovprior{t}(\rnnParam)\right)$ for $l=1, \ldots, L$. The parameters of the Gaussian prior are obtained using $\text{RNN}(\cdot; \rnnParam)$ as per \eqref{eq:RNNPrior}. Crucially, to enable gradient computation in the sampling step of \eqref{eq:marginal-dist-LB}, we use the reparameterization trick for multivariate Gaussian \cite{kingma2013auto} as in \eqref{eq:inferencetimesampling}. 
%\begin{IEEEeqnarray}{rl}
%\bstate^{(l)}_{t}\left(\rnnParam\right) &\sim \mathcal{N}\left(\bstate_{t}; \stateMeanprior{t}(\rnnParam), \stateCovprior{t}(\rnnParam)\right) \nonumber\\
%&= \stateMeanprior{t}(\rnnParam) + \stateCovprior{t}^{{1}/{2}}(\rnnParam) \boldsymbol{\epsilon}^{(l)}, \nonumber \\
%\label{eq:reparameterization-trick}
%\end{IEEEeqnarray}
%where $\boldsymbol{\epsilon}^{(l)} \sim \normaldist{\boldsymbol{\epsilon}}{\boldsymbol{0}}{\mathbf{I}}$ for $l=1, \ldots, L$ and $\stateCovprior{t}^{{1}/{2}}(\rnnParam)$ is computed using Cholesky decomposition of $\stateCovprior{t}(\rnnParam)$. 
This is also schematically shown in Fig. \ref{fig:reparam_trick}.
%\begin{color}{blue}
Next, we consider the joint posterior distribution term in \eqref{eq:semi-supervised-learning-problem}. Following the standard approach of mean-field approximation \cite[Chap. 10]{bishop2006pattern} and maintaining causality \cite{ghosh2025pdanse}, \cite[Section II-C]{ghosh2024data}, we assume the following decomposition of the joint posterior distribution
\begin{IEEEeqnarray}{rl}
\log p\left(\bstate_{1:T} \vert \bmeas_{1:T}; \rnnParam\right) &= \log \prod_{t=1}^{T} p\left(\bstate_{t} \vert  \bstate_{1:t-1} , \bmeas_{1:T}; \rnnParam\right)
\label{eq:jointposteriordecomposition_initial}
\end{IEEEeqnarray}
Under the mean-field approximation, we can simplify the RHS of \eqref{eq:jointposteriordecomposition_initial} as follows
\begin{IEEEeqnarray}{rl}
\log \prod_{t=1}^{T} p\left(\bstate_{t} \vert  \bstate_{1:t-1} , \bmeas_{1:T}; \rnnParam\right)
&= \log \prod_{t=1}^{T} p\left(\bstate_{t} \vert \bmeas_{1:T}; \rnnParam\right) 
\label{eq:jointposteriordecomposition_meanfield}
\end{IEEEeqnarray}
Then under the causality assumption, we have the following equality
\begin{IEEEeqnarray}{rl}
\log \prod_{t=1}^{T} p\left(\bstate_{t} \vert \bmeas_{1:T}; \rnnParam\right) &= \log \prod_{t=1}^{T} p\left(\bstate_{t} \vert \bmeas_{1:t}; \rnnParam\right).
\label{eq:jointposteriordecomposition_causality}
\end{IEEEeqnarray}
Combining \eqref{eq:jointposteriordecomposition_meanfield} and \eqref{eq:jointposteriordecomposition_causality}, we have
\begin{IEEEeqnarray}{rl}
\log p\left(\bstate_{1:T} \vert \bmeas_{1:T}; \rnnParam\right) &= \log \prod_{t=1}^{T} p\left(\bstate_{t} \vert \bmeas_{1:t}; \rnnParam\right) \nonumber \\
&= \sum_{t=1}^{T} \log  p\left(\bstate_{t} \vert \bmeas_{1:t}; \rnnParam\right).
\label{eq:jointposteriordecomposition}
\end{IEEEeqnarray}
%\end{color}
We can employ Bayes' rule on the component term of the RHS of \eqref{eq:jointposteriordecomposition} as follows 
\begin{IEEEeqnarray}{rl}
\log  p\left(\bstate_{t} \vert \bmeas_{1:t}; \rnnParam\right)  &= \log  \left(\frac{p\left(\bmeas_{t} \vert \bstate_{t} \right)p\left(\bstate_{t} \vert \bmeas_{1:t-1}; \rnnParam \right)}{p\left(\bmeas_{t} \vert \bmeas_{1:t-1}; \rnnParam \right)} \right)\nonumber \\
&= \log p\left(\bmeas_{t} \vert \bstate_{t} \right) + \log p\left(\bstate_{t} \vert \bmeas_{1:t-1}; \rnnParam \right) \nonumber \\
&- \log p\left(\bmeas_{t} \vert \bmeas_{1:t-1}; \rnnParam \right).
\label{eq:marginalposteriorBayes}
\end{IEEEeqnarray}
Using \eqref{eq:jointposteriordecomposition}, \eqref{eq:marginalposteriorBayes}, \eqref{eq:marginal-dist-decomposition} and \eqref{eq:marginal-dist-LB}, we simplify the optimization problem in \eqref{eq:semi-supervised-learning-problem} to
\begin{IEEEeqnarray}{rl}
\label{eq:semi-supervised-learning-problem2}
     \rnnParam^\star 
    &= \arg\max\limits_{\rnnParam} \log p\left(\left\lbrace \left(\bstate^{(j)}_{1:T} \Big\vert \bmeas^{(j)}_{1:T}\right)\right\rbrace_{j \in \IndexSet_s}; \rnnParam\right) \nonumber\\
    &+ \log p\left(\left\lbrace \left(\bmeas^{(i)}_{1:T}\right)\right\rbrace_{i \in \IndexSet_u \cup \IndexSet_s}; \rnnParam\right)\nonumber\\
    %&= \arg\max\limits_{\rnnParam} \sum_{j \in \IndexSet_s} \sum_{t=1}^{T} \log p\left(\bstate^{(j)}_{t} \Big\vert \bmeas^{(j)}_{1:t}; \rnnParam\right) \nonumber\\
    %&+ \sum_{i \in \IndexSet_s \cup \IndexSet_u} \sum_{t=1}^{T}\log  p\left(\bmeas_t^{(i)} \vert \bmeas^{(i)}_{1:t-1} ; \rnnParam\right) \nonumber\\
    &= \arg\max\limits_{\rnnParam} \sum_{j \in \IndexSet_s} \sum_{t=1}^{T} \Big\lbrace \log p\left(\bmeas_t^{(j)} \vert \bstate_t^{(j)}\right)\nonumber \\
    &+ \log p\left(\bstate_t^{(j)} \vert \bmeas^{(j)}_{1:t-1}; \pmb{\theta}\right) 
 \sminus \log p\left(\bmeas^{(j)}_t \vert \bmeas^{(j)}_{1:t-1} ; \rnnParam\right)\Big\rbrace \nonumber \\ 
    &+ \sum_{i \in \IndexSet_s \cup \IndexSet_u} \sum_{t=1}^{T}\log  p\left(\bmeas_t^{(i)} \vert \bmeas^{(i)}_{1:t-1} ; \rnnParam\right) \nonumber  \\
    &= \arg\max\limits_{\rnnParam} \sum_{j \in \IndexSet_s} \sum_{t=1}^{T} \Big\lbrace \log p\left(\bmeas_t^{(j)} \vert \bstate_t^{(j)}\right) \nonumber\\
    &+ \log p\left(\bstate_t^{(j)} \vert \bmeas^{(j)}_{1:t-1}; \pmb{\theta}\right) \!\!\Big\rbrace + \sum_{i \in \IndexSet_u} \sum_{t=1}^{T}\log  p\left(\bmeas_t^{(i)} \vert \bmeas^{(i)}_{1:t-1} ; \rnnParam\right) \nonumber \\
    &\equiv \arg\min\limits_{\rnnParam} \Big\lbrace  \LossSup\left(\Dataset_s; \rnnParam\right) 
    + \LossUnsup\left( \Dataset_u; \rnnParam\right)\Big\rbrace. 
    %&= \arg\max\limits_{\rnnParam} \Big\lbrace \sum_{j \in \IndexSet_s} \sum_{t=1}^{T} \Big\lbrace \log \mathcal{N}\!\left(\bmeas_t; \bhn\left(\bstate_t\right), \bmnoiseCov \right) \\
    %&+ \log \mathcal{N}\!\left(\bstate_t; \stateMeanprior{t}\left(\rnnParam\right), \stateCovprior{t}\left(\rnnParam\right)\right) \Big\rbrace  \\
    %&+ \sum_{i \in \IndexSet_u} \sum_{t=1}^{T} \frac{1}{L} \sum\limits_{l=1}^{L}\log \mathcal{N}\!\left(\bmeas_t; \bhn\left(\bstate_t^{(l)}\right), \bmnoiseCov \right) \Big\rbrace \\ 
\end{IEEEeqnarray}
We note from the last line in \eqref{eq:semi-supervised-learning-problem2} that the optimization involves a combination of a supervised and an unsupervised loss. The supervised loss term is 
\begin{IEEEeqnarray}{RL}
    \LossSup\left(\Dataset_{s}; \rnnParam\right) 
    %\LossSup\left(\Dataset_s; \rnnParam\right) \
    &= -\sum_{j \in \IndexSet_s} \sum_{t=1}^{T} \Big\lbrace \log p\left(\bmeas_t^{(j)} \vert \bstate_t^{(j)}\right) \nonumber\\
    &+ \log p\left(\bstate_t^{(j)} \vert \bmeas^{(j)}_{1:t-1}; \pmb{\theta}\right) \!\!\Big\rbrace \nonumber \\
    &= -\sum_{j \in \IndexSet_s} \sum_{t=1}^{T} \Big\lbrace\!\!\log \mathcal{N}\!\left(\bmeas_t^{(j)}; \bhn\left(\bstate_t^{(j)}\right), \bmnoiseCov \right) \nonumber\\
    &+ \log \mathcal{N}\!\left(\bstate_t^{(j)}; \stateMeanprior{t}^{(j)}\left(\rnnParam\right), \stateCovprior{t}^{(j)}\left(\rnnParam\right)\right) \!\!\!\Big\rbrace, 
\label{eq:suplossnonlinear}
\end{IEEEeqnarray}
where $\lbrace \stateMeanprior{t}^{(j)}\left(\rnnParam\right), \stateCovprior{t}^{(j)}\left(\rnnParam\right)\rbrace$ is obtained using \eqref{eq:RNNPrior} with $\bmeas^{(j)}_{1:t-1}$ as input $(\text{ s.t. } j \in \IndexSet_s)$. The unsupervised loss term using the lower bound \eqref{eq:marginal-dist-LB} is 
\begin{IEEEeqnarray}{rl}
    &\LossUnsup\left( \Dataset_u; \rnnParam\right) \nonumber\\
%    &= -\sum_{i \in \IndexSet_u} \sum_{t=1}^{T}\log  p\left(\bmeas_t^{(i)} \vert \bmeas^{(i)}_{1:t-1} ; \rnnParam\right) \nonumber \\
    &= -\frac{1}{L}\sum_{i \in \IndexSet_u}\sum_{t=1}^{T} \sum_{l=1}^{L}{\log \mathcal{N}\!\left(\bmeas_t^{(i)}; \bhn\!\left(\bstate_t^{(i,l)}\left(\rnnParam\right)\!\right), \bmnoiseCov \right)}.
\label{eq:unsuplossnonlinear}
\end{IEEEeqnarray}
Here $\bstate_t^{(i,l)}\left(\rnnParam\right)$ is sampled for $l=1, \ldots, L$ using \eqref{eq:inferencetimesampling}, with $\lbrace \stateMeanprior{t}^{(i)}\left(\rnnParam\right), \stateCovprior{t}^{(i)}\left(\rnnParam\right)\rbrace$ 
obtained using \eqref{eq:RNNPrior} with $\bmeas^{(i)}_{1:t-1}$ as input $(\text{ s.t. } i \in \IndexSet_u)$. We remark that the effect of the supervised loss term \eqref{eq:suplossnonlinear} on the optimization problem \eqref{eq:semi-supervised-learning-problem2} depends on the number of samples available in $\Dataset_{s}$, i.e. $\vert \IndexSet_{s} \vert$. Typically $ \vert \IndexSet_{s} \vert \ll N$ in a semi-supervised learning scenario. In the absence of any supervised data, i.e. $\vert \IndexSet_{s} \vert = 0$, we obtain an unsupervised learning scenario from $\eqref{eq:semi-supervised-learning-problem2}$ as a subcase, as follows:
\begin{IEEEeqnarray}{rl}
\rnnParam^\star &= \arg\max\limits_{\rnnParam} \log p\left({\Dataset_{\text{semi}}}; \rnnParam\right) \Big \vert_{\vert \IndexSet_{s} \vert = 0} \nonumber\\
&=\arg\max\limits_{\rnnParam} \log p\left({\Dataset_{{u}}}; \rnnParam\right) \nonumber\\
%&=\arg\max\limits_{\rnnParam} \log p\left(\left\lbrace \left(\bstate_{1:T}^{(j)}, \bmeas_{1:T}^{(j)}\right)\right\rbrace_{j \in \IndexSet_{s}}, \left\lbrace \bmeas_{1:T}^{(i)}\right\rbrace_{i \in \IndexSet_{u}}; \rnnParam\right) \nonumber \\
&=\arg\max\limits_{\rnnParam}  \log p\left(\left\lbrace \bmeas_{1:T}^{(i)}\right\rbrace_{i \in \IndexSet_{s}\cup\IndexSet_{u}}; \rnnParam\right)\nonumber \\
&= \arg\max\limits_{\rnnParam} \sum_{i \in \IndexSet_u} \sum_{t=1}^{T}\log  p\left(\bmeas_t^{(i)} \vert \bmeas^{(i)}_{1:t-1} ; \rnnParam\right) \nonumber \\
&\equiv \arg\min\limits_{\rnnParam} \LossUnsup\left( \Dataset_u; \rnnParam\right),
\label{eq:unsupervised-learning-problem}
\end{IEEEeqnarray}
where $ \LossUnsup\left( \Dataset_u; \rnnParam\right)$ is defined in \eqref{eq:unsuplossnonlinear}. Note that in the case of DANSE, we would be able to directly optimize $\log  p\left(\bmeas_t \vert \bmeas_{1:t-1}; \rnnParam\right)$ as it is possible to obtain it in closed-form as shown in \eqref{eq:pyt_given_prev}. However, in the presence of a nonlinear measurement function $\bhn$, we resort to maximizing a lower bound of $\log  p\left(\bmeas_t \vert \bmeas_{1:t-1}; \rnnParam\right)$ given in \eqref{eq:marginal-dist-LB}, as shown in \eqref{eq:unsupervised-learning-problem}. 

%\begin{color}{blue}
\subsection{Supervised version of pDANSE}\label{sec:pdanse_supervised}
Recall the supervised loss term for pDANSE defined in \eqref{eq:suplossnonlinear} of the revised version of the manuscript as 
\begin{equation}\label{eq:suplossnonlinear_pdanse}
\begin{array}{rl}
    \LossSup\left(\Dataset_{s}; \rnnParam\right) 
    %\LossSup\left(\Dataset_s; \rnnParam\right) \
    &= -\sum_{j \in \IndexSet_s} \sum_{t=1}^{T} \Big\lbrace \log p\left(\bmeas_t^{(j)} \vert \bstate_t^{(j)}\right) \nonumber\\
    &+ \log p\left(\bstate_t^{(j)} \vert \bmeas^{(j)}_{1:t-1}; \pmb{\theta}\right) \!\!\Big\rbrace \nonumber \\
    &= -\sum_{j \in \IndexSet_s} \sum_{t=1}^{T} \Big\lbrace\!\!\log \mathcal{N}\!\left(\bmeas_t^{(j)}; \bhn\left(\bstate_t^{(j)}\right), \bmnoiseCov \right) \nonumber\\
    &+ \log \mathcal{N}\!\left(\bstate_t^{(j)}; \stateMeanprior{t}^{(j)}\left(\rnnParam\right), \stateCovprior{t}^{(j)}\left(\rnnParam\right)\right) \!\!\!\Big\rbrace, 
\end{array}
\end{equation}
This loss term depends only on supervised data $\Dataset_{s} = \left\lbrace \left(\bstate_{1:T}^{(j)}, \bmeas_{1:T}^{(j)} \right)\right\rbrace_{j \in \IndexSet_{s}}$. In fact, at a given time point $t$, only $\bmeas_{1:t}$ and $\bstate_t$ are used in the loss calculation. Following this observation, we recognize that the use of only the loss term in \eqref{eq:suplossnonlinear_pdanse} in the optimization problem leads to a version of pDANSE depending only on $\Dataset_s$. The inference procedure would be the same as the proposed semi-supervised pDANSE. Effectively the algorithm tries to map the $\bmeas_{1:t}$ to $\bstate_t$ by solving the following supervised learning-based optimization problem
\begin{equation}\label{eq:pdanse-supervised-optimization}
    \min\limits_{\rnnParam}  \LossSup\left(\Dataset_s; \rnnParam\right).  
\end{equation}
We refer to this version of pDANSE optimized using \eqref{eq:pdanse-supervised-optimization} as pDANSE-Supervised in the subsequent experiments.
%\end{color}
\section{Experiments and Results} \label{sec:experiments_and_results}
In this section, we show the BSE performance of the proposed $\pdanse$ against the PF. pDANSE does not use any knowledge of STM, and PF uses the full knowledge of STM. {Also, PF is used as a model-driven benchmark.} {We use a bootstrap PF in our experiments}.  We investigate four nonlinear measurement systems. They are: cubic nonlinearity, camera model nonlinearity \cite{buchnik2023latent}, half-wave rectification nonlinearity, and Cartesian-to-spherical nonlinearity \cite{li2001survey}. ⁄{For our experiments, we use a stochastic Lorenz-$63$ system} \cite{lorenz1963deterministic} as a benchmark process, and it has been extensively used in the literature \cite{revach2022unsupervised, ghosh2023dansejrnl, garcia2019combining, revach2022kalmannet, ni2024adaptive}. The Lorenz-$63$ STM is 
\begin{IEEEeqnarray}{RL}
     \bstate_{t+1} 
     &= \mathbf{F}_{t}(\bstate_{t})\bstate_{t} + \bpnoise_t \in \setR^3, \nonumber \\
     \text{s.t. }\mathbf{F}_{t}(\bstate_{t}) &= \matrixexp{\left(
    \begin{bmatrix}
        -10 & 10 & 0 \\
        28 & -1 & -\state_{t, 1} \\
        0 & \state_{t, 1} & -\frac{8}{3} \\ 
    \end{bmatrix}\Delta 
    \right)}, 
    \label{eq:lorenz63nlssm}
\end{IEEEeqnarray}
%\begin{IEEEeqnarray}{rl}\label{eq:lorenz_mexp}
%    \text{s.t. }\mathbf{F}_{t}(\bstate_{t}) &= \matrixexp{\left(
%    \begin{bmatrix}
%        -10 & 10 & 0 \\
%        28 & -1 & -\state_{t, 1} \\
%        0 & \state_{t, 1} & -\frac{8}{3} \\ 
%    \end{bmatrix}\Delta 
%    \right)}, 
%\end{IEEEeqnarray}
%{Can not we write `exp' instead of `e'? `e' is used for process noise.}
where the process noise $\bpnoise_t \sim \normaldist{\bpnoise_t}{\boldsymbol{0}}{\bpnoiseCov}$ with $\bpnoiseCov=\sigmapnoise^2 \mathbf{I}_{3}$, the step-size $\Delta=0.02 \text{ seconds}$, and $\sigmapnoise^2$ was set corresponding to $-10$ dB. In our simulations, we use $5$'$\text{th}$ order Taylor series approximation for $\mathbf{F}_{t}(\bstate_{t})$ shown in \eqref{eq:lorenz63nlssm} similar to \cite{revach2022kalmannet, ghosh2023dansejrnl, garcia2019combining}. {Additionally, we also show the performance of pDANSE for the stochastic Lorenz-$96$ system with a half-wave, rectified measurement system.} %Note that $\mathrm{dim}(\mathbf{x}_t) \triangleq m = 3$ for all our experiments.

For ease of illustration, we define $\kappa \triangleq N_s / N$ to indicate the $\%$ of labelled data usage in training of $\pdanse$. Note that $0 \leq \kappa \leq 1$; $\kappa=0\,\,(\text{ or }0\%)$ refers to unsupervised learning and $\kappa=1\,\,(\text{ or }100\%)$ refers to a fully supervised learning framework. We train $\pdanse$ on $\mathcal{D}_{\text{semi}}$ having a specified $\kappa$, $N=N_s + N_u=1000$ trajectories, with each trajectory having the same length $T=200$. After training, we test $\pdanse$ and the PF 
on a labelled dataset $\Dataset_{\text{test}} = \Big\lbrace  \left(\bstate_{1:T_{\text{test}}}^{(j)}, \bmeas_{1:T_{\text{test}}}^{(j)} \right) \Big\rbrace_{j=1}^{N_{\text{test}}}$ having $N_{\text{test}}=100$ trajectories each of length $T_{\text{test}}=2000$, similar to \cite{ghosh2023dansejrnl}.  %We used $L=10$ MC samples for computing the posterior moments using \eqref{eq:approxposteriormean}, \eqref{eq:approxposteriorcov}.
For our experiments, we generate noisy test data where we use $\bmnoiseCov = \sigma^2_{\mnoise} \mathbf{I}_{3}$; here $\sigma^2_{\mnoise}$ is chosen using the average signal-to-measurement noise ratio (SMNR) in dB scale  \cite[section III-C]{ghosh2023dansejrnl}  
\begin{IEEEeqnarray*}{RL}\label{eq:SMNR}
    &\mathrm{SMNR} \\
    &=  \frac{1}{N_{\text{test}}} \sum \limits_{j=1}^{N_{\text{test}}}  \!\! 10 \log_{10}  \sum_{t=1}^{T_{\text{test}}^{(j)}}\frac{ \mathbb{E}\left[  \Big\Vert \bhn \left(\bstate_t^{(j)}\right) \sminus \mathbb{E}\left[ \bhn \left(\bstate_t^{(j)}\right) \right] \Big\Vert_2^2 \right]}{n \sigma_w^2}.  
    %\!\!&= \!\! \frac{1}{N_{\text{test}}} \!\! \sum\limits_{j=1}^{N_{\text{test}}} \!\! 10 \log_{10} \!\! \left( \sum_{t=1}^{T_{\text{test}}^{(j)}}\frac{\mathbb{E}\lbrace  \| \bmeasmat\bstate_t^{(j)} - \mathbb{E}\lbrace \bmeasmat \bstate_t^{(j)} \rbrace \|_2^2 \rbrace}{n\sigma_w^2}\right).
\end{IEEEeqnarray*}
A low value of SMNR indicates a high strength of the measurement noise. Similarly, noisy measurements of the training dataset are generated.

The BSE performance evaluation metric is the normalized mean squared error (NMSE) in dB scale, defined as \cite[section III]{ghosh2023dansejrnl}
\begin{IEEEeqnarray*}{RL}
    \mathrm{NMSE} &= \frac{1}{N_{\text{test}}} \sum_{j=1}^{N_{\text{test}}} 10\log_{10}\left(\frac{\sum_{t=1}^{T_{\text{test}}^{(j)}}\| \bstate_t^{(j)} \sminus \hat{\bstate}_t^{(j)} \|_2^2}{\sum_{t=1}^{T_{\text{test}}^{(j)}} \| \bstate_t^{(j)}\|_2^2}\right).
\end{IEEEeqnarray*}
In the above equation, $\hat{\bstate}_t$ is set as the posterior mean estimate $\stateMeanposterior{t}\left(\rnnParam\right)$, shown in  \eqref{eq:approxposteriormean}.
%For comparison, we use a model-driven PF using $100$ particles.  

%\noindent \textbf{Training details}: 
%Our interest is primarily the performance of the semi-supervised trained $\pdanse$, i.e. when $0 < \kappa < 1$. Note that unsupervised / supervised versions of $\pdanse$ can be obtained with $\kappa = 0 \text{ or } 1$ respectively. In the subsequent section, we refer to the unsupervised version of $\pdanse$ as $\pdanse \left(\kappa=0\%\right)$ and compare against the semi-supervised version of $\pdanse$ corresponding to the setting of $\kappa=2\%$ in $\Dataset_{\text{semi}}$. 

The RNN that we used in the implementation is a GRU, with $2$ hidden layers and $80$ hidden units in each layer \cite{choPropertiesNeuralMachine2014}. %This is to due to their simplified structure compared to long-short-term-memory networks (LSTMs) \cite{hochreiter1997long}. 
The output of the GRU was further mapped using $1$-layer, feed-forward networks with $128$ hidden units to obtain the mean and covariance in \eqref{eq:RNNPrior} similar to \cite{ghosh2024data}. The specific GRU architecture was decided by grid search on a held-out validation dataset. We perform the simulations using Python and PyTorch \cite{paszke2019pytorch} \footnote{The implementation code will be made available upon request.}. $\pdanse$ was trained with a single GPU support using mini-batch gradient descent with a batch size of $128$. The optimizer was Adam with an adaptive learning rate starting at $10^{-4}$ and reduced by $20\%$ after a fixed number of epochs \cite{kingma2014adam}.

\begin{figure}[t]
    \centering
    \includegraphics[width=\linewidth]{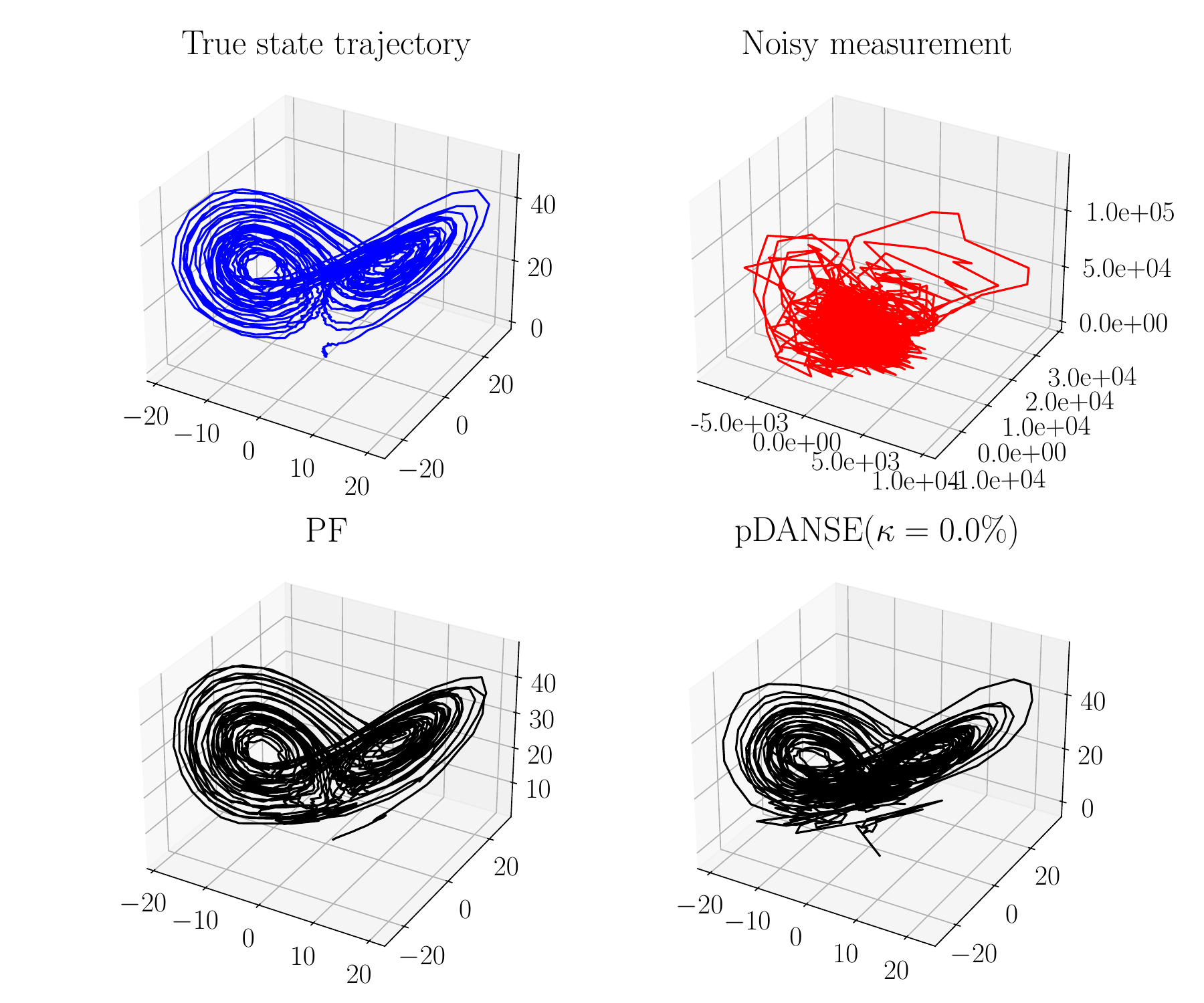}
    \caption{Visual illustration of the BSE performance for cubic nonlinearity in an unsupervised learning setup at $\text{SMNR} = 20 \text{ dB}$. Top-left - A true state trajectory for the stochastic Lorenz-$63$ system from $\Dataset_{\text{test}}$. Top-right - The corresponding noisy measurement trajectory. Bottom-left - The estimated trajectory of PF. Bottom-right- The estimated trajectory of pDANSE. %Demonstrating the performance of $\pdanse$ using $\kappa = 0.0$ for the BSE problem using noisy, element-wise cubic measurements of the Lorenz-$63$ process, at $\text{SMNR} = 20 \text{ dB}$. The estimated state trajectories (shown in black) are the corresponding posterior mean estimates using PF and $\pdanse$ $(\kappa=0\%)$. The PF is model-driven and knows the exact state dynamics.
    }
    \label{fig:pdanse_cubic_3d}
\end{figure}
\begin{figure}[t]
    \centering
    \input{figs/journalpaper/cubic/LoreznSSM_Axiswise_pdanse_nsup_pf_smnr20.0dB_plot}
    \caption{Demonstrating the qualitative performance of $\pdanse$ ($\kappa= 0\%$) for the BSE problem using noisy, element-wise cubic measurements of the Lorenz-$63$ process, at $\text{SMNR} = 20 \text{ dB}$. A chosen realization of a true state trajectory from $\Dataset_{\text{test}}$ is shown in black. The estimated state trajectories are the corresponding posterior mean estimates using PF (shown in blue) and $\pdanse$ (shown in green). The shaded region indicates a $\pm 3 \sigma$ confidence for the posterior mean estimate. }
    \label{fig:pdanse_cubic_3d_axiswise}
\end{figure}
\begin{figure}[t]
    \centering
    \scalebox{0.97}{% This file was created with tikzplotlib v0.10.1.
\begin{tikzpicture}

\definecolor{darkgray176}{RGB}{176,176,176}
\definecolor{lightgray204}{RGB}{204,204,204}

\begin{axis}[
legend cell align={left},
legend style={fill opacity=0.8, draw opacity=1, text opacity=1, draw=lightgray204},
tick align=outside,
tick pos=left,
x grid style={darkgray176},
xlabel={SMNR (in dB)},
xmajorgrids,
xmin=-1.5, xmax=31.5,
xtick style={color=black},
xtick={-5,0,5,10,15,20,25,30,35},
xticklabels={
  \(\displaystyle {\ensuremath{-}5}\),
  \(\displaystyle {0}\),
  \(\displaystyle {5}\),
  \(\displaystyle {10}\),
  \(\displaystyle {15}\),
  \(\displaystyle {20}\),
  \(\displaystyle {25}\),
  \(\displaystyle {30}\),
  \(\displaystyle {35}\)
},
y grid style={darkgray176},
ylabel={NMSE (in dB)},
ymajorgrids,
ymin=-26.0066152870655, ymax=-4.20689558386803,
ytick style={color=black},
ytick={-27.5,-25,-22.5,-20,-17.5,-15,-12.5,-10,-7.5,-5,-2.5},
yticklabels={
  \(\displaystyle {\ensuremath{-}27.5}\),
  \(\displaystyle {\ensuremath{-}25.0}\),
  \(\displaystyle {\ensuremath{-}22.5}\),
  \(\displaystyle {\ensuremath{-}20.0}\),
  \(\displaystyle {\ensuremath{-}17.5}\),
  \(\displaystyle {\ensuremath{-}15.0}\),
  \(\displaystyle {\ensuremath{-}12.5}\),
  \(\displaystyle {\ensuremath{-}10.0}\),
  \(\displaystyle {\ensuremath{-}7.5}\),
  \(\displaystyle {\ensuremath{-}5.0}\),
  \(\displaystyle {\ensuremath{-}2.5}\)
},
width=\linewidth,
height=0.75\linewidth
]
\path [draw=red, semithick]
(axis cs:0,-11.4952583312988)
--(axis cs:0,-8.67460632324219);

\path [draw=red, semithick]
(axis cs:10,-17.2666368484497)
--(axis cs:10,-11.5848093032837);

\path [draw=red, semithick]
(axis cs:20,-21.1389627456665)
--(axis cs:20,-13.9091596603394);

\path [draw=red, semithick]
(axis cs:30,-25.0157189369202)
--(axis cs:30,-15.2045874595642);

\path [draw=blue, semithick]
(axis cs:0,-5.68764507770538)
--(axis cs:0,-5.19779193401337);

\path [draw=blue, semithick]
(axis cs:10,-8.446009516716)
--(axis cs:10,-7.47339737415314);

\path [draw=blue, semithick]
(axis cs:20,-16.1877462863922)
--(axis cs:20,-14.7637422084808);

\path [draw=blue, semithick]
(axis cs:30,-20.7465742230415)
--(axis cs:30,-19.2584039568901);

\addplot [thick, red, mark=diamond, mark size=4, mark options={solid,fill opacity=0}]
table {%
0 -10.0849323272705
10 -14.4257230758667
20 -17.5240612030029
30 -20.1101531982422
};
\addlegendentry{$\text{PF}$}
\addplot [thick, blue, mark=o, mark size=4, mark options={solid,fill opacity=0}]
table {%
0 -5.44271850585938
10 -7.95970344543457
20 -15.4757442474365
30 -20.0024890899658
};
\addlegendentry{$\text{pDANSE}  (\kappa=0\%)$}
\end{axis}

\end{tikzpicture}}
    \caption{NMSE (in dB) on $\Dataset_{\text{test}}$ vs. SMNR (in dB), demonstrating the performance of $\pdanse$ ($\kappa=0\%$) for the BSE task using noisy, cubic measurements of the Lorenz-$63$ process. The nonlinear function $\bhn$ is defined in \eqref{eq:cubic_measurement_fn}. $\pdanse$ was trained using $N=1000, T=200$.}
    \label{fig:nmse_cubic_pdanse}
\end{figure}
\begin{figure*}[t]
    \centering
    \includegraphics[width=\linewidth]{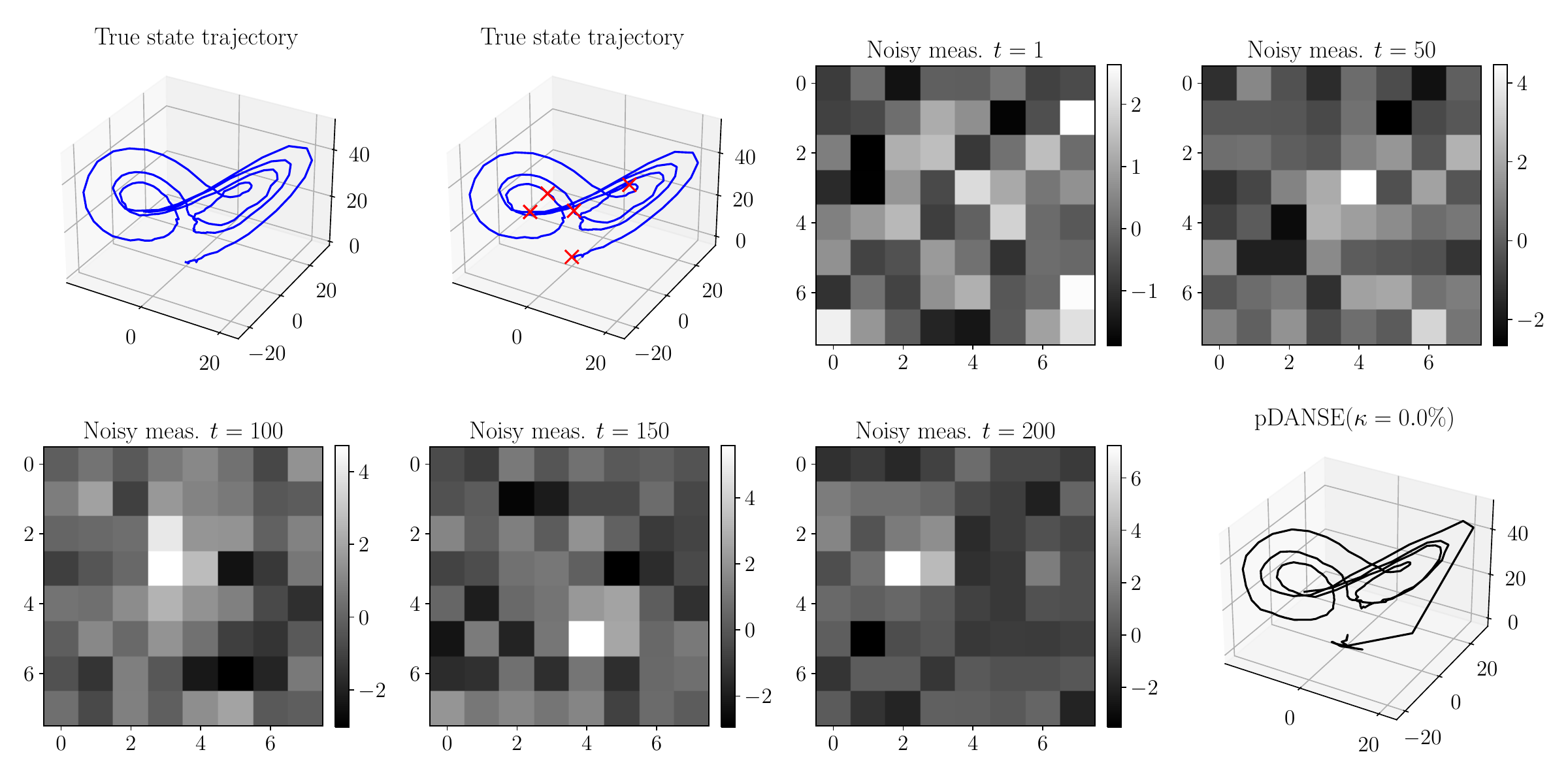}
    \caption{Visual illustration of the BSE performance for camera model nonlinearity ($8 \times 8$ image) in an unsupervised learning setup, at $\text{SMNR} = 0 \text{ dB}$. A true state trajectory from $\Dataset_{\text{test}}$ is shown in blue (first row, first plot). The snapshots of the noisy measurements at five different time instants ($t=1,50,100,150, 200$) marked by the red crosses in the true state trajectory (first row, second plot) are shown as images (first row, third image to second row, third image). The estimated state trajectory using $\pdanse$ is shown in black (second row, last plot). A shorter test set trajectory ($T_{\text{test}} = 200$) is shown for clarity.}
    \label{fig:meas-highdim-reconstructed}
\end{figure*}

%\subsection{Results}\label{sec:results} 
\subsection{Experiments in unsupervised learning setups}\label{sec:results_unsupervised_learning}
We begin our experiments with two nonlinearities in unsupervised learning setups where $\kappa = 0\%$. The nonlinearities are: an element-wise, cubic nonlinearity and a camera model nonlinearity inspired by \cite{buchnik2023latent}.  

\subsubsection{Cubic nonlinearity}
Following \eqref{eq:measurementsys}, the elements of $\bhn \left(\bstate_{t}\right)$ are as follows:
\begin{IEEEeqnarray}{rl}\label{eq:cubic_measurement_fn}
    h_{i}\left(\bstate_t\right) = \state^{3}_{t,i}, \, \, \, i=1, 2, 3.
\label{eq:lorenzssm_hfn}
\end{IEEEeqnarray}
%The measurement noise $\bmnoise_{t}$ is i.i.d., additive, white Gaussian noise as in \eqref{eq:measurementsys}, s.t. $\bmnoise_{t} \sim \mathcal{N}\left(\bmnoise_{t}; \boldsymbol{0}, \bmnoiseCov\right)$.
Subsequently, we added Gaussian measurement noise $\bmnoise_{t} \in \setR^{3}$, s.t. $\bmnoise_{t} \sim \normaldist{\bmnoise_{t}}{\boldsymbol{0}}{\sigma^{2}_{\mnoise} \mathbf{I}_{3}}$ to create noisy measurements $\bmeas_{t} \in \setR^{3}$. 
For $\pdanse$, we use $L=10$ MC samples for training and testing. We compare our approach with PF having $300$ particles. A higher number of particles was used for the PF to serve as a strong benchmark. %Both PF and UKF know the true underlying state dynamics. 
In Fig. \ref{fig:pdanse_cubic_3d}, we provide a visual illustration of the performance of $\pdanse$, and compare against that of the PF, at $\text{SMNR}=20 \text{ dB}$. We observe that $\pdanse$ provides good performance despite being trained in an unsupervised manner. In Fig. \ref{fig:pdanse_cubic_3d_axiswise}, we show a time plot of a randomly selected true state trajectory and corresponding estimates using PF and $\pdanse$, along with $\pm 1 \sigma$ confidence regions. Finally, we show NMSE-versus-SMNR performance comparison vis-a-vis PF in Fig. \ref{fig:nmse_cubic_pdanse}. {We observe that the gap between pDANSE and PF is higher at the low range of SMNR.} %Note that PF is an asymptotically optimal estimator. 
%\begin{figure}[t]
%    \centering
%    \includegraphics[width=\linewidth]{figs/journalpaper/relu/LoreznSSM_pdanse_nsup20_pf_smnr10.0dB_3x2_plot.pdf}
%    \caption{Demonstrating the performance of $\pdanse$ for different $\kappa$ for the BSE problem using noisy, element-wise rectified measurements of the Lorenz-$63$ process, at $\text{SMNR} = 10 \text{ dB}$. The estimated state trajectories (shown in black) are the corresponding posterior mean estimates using PF and $\pdanse$ at different $\kappa$ values. PF is model-driven and knows the exact state dynamics.}
%    \label{fig:pdanse_relu_3d}
%\end{figure}

\begin{figure}[t]
    \centering
    \scalebox{1.0}{% This file was created with tikzplotlib v0.10.1.
\begin{tikzpicture}

\definecolor{darkgray176}{RGB}{176,176,176}
\definecolor{lightgray204}{RGB}{204,204,204}

\begin{axis}[
legend cell align={left},
legend style={fill opacity=0.8, draw opacity=1, text opacity=1, draw=lightgray204},
tick align=outside,
tick pos=left,
x grid style={darkgray176},
xlabel={SMNR (in dB)},
xmajorgrids,
xmin=-1.5, xmax=31.5,
xtick style={color=black},
xtick={-5,0,5,10,15,20,25,30,35},
xticklabels={
  \(\displaystyle {\ensuremath{-}5}\),
  \(\displaystyle {0}\),
  \(\displaystyle {5}\),
  \(\displaystyle {10}\),
  \(\displaystyle {15}\),
  \(\displaystyle {20}\),
  \(\displaystyle {25}\),
  \(\displaystyle {30}\),
  \(\displaystyle {35}\)
},
y grid style={darkgray176},
ylabel={NMSE (in dB)},
ymajorgrids,
ymin=-39.831679058075, ymax=-20.6378930091858,
ytick style={color=black},
ytick={-40,-37.5,-35,-32.5,-30,-27.5,-25,-22.5,-20},
yticklabels={
  \(\displaystyle {\ensuremath{-}40.0}\),
  \(\displaystyle {\ensuremath{-}37.5}\),
  \(\displaystyle {\ensuremath{-}35.0}\),
  \(\displaystyle {\ensuremath{-}32.5}\),
  \(\displaystyle {\ensuremath{-}30.0}\),
  \(\displaystyle {\ensuremath{-}27.5}\),
  \(\displaystyle {\ensuremath{-}25.0}\),
  \(\displaystyle {\ensuremath{-}22.5}\),
  \(\displaystyle {\ensuremath{-}20.0}\)
},
width=0.95\linewidth,
height=0.75\linewidth
]
\path [draw=red, semithick]
(axis cs:0,-29.5838632583618)
--(axis cs:0,-22.6492681503296);

\path [draw=red, semithick]
(axis cs:5,-32.247697353363)
--(axis cs:5,-23.1633629798889);

\path [draw=red, semithick]
(axis cs:10,-33.982141494751)
--(axis cs:10,-25.7866611480713);

\path [draw=red, semithick]
(axis cs:20,-37.1116361618042)
--(axis cs:20,-27.3764009475708);

\path [draw=red, semithick]
(axis cs:30,-38.9592342376709)
--(axis cs:30,-25.9434833526611);

\path [draw=blue, semithick]
(axis cs:0,-23.363353729248)
--(axis cs:0,-21.5103378295898);

\path [draw=blue, semithick]
(axis cs:5,-27.4162261486053)
--(axis cs:5,-24.9433853626251);

\path [draw=blue, semithick]
(axis cs:10,-29.475123167038)
--(axis cs:10,-26.4581005573273);

\path [draw=blue, semithick]
(axis cs:20,-31.6195483207703)
--(axis cs:20,-28.1813893318176);

\path [draw=blue, semithick]
(axis cs:30,-31.3798639774323)
--(axis cs:30,-28.7900426387787);

\addplot [thick, red, mark=diamond, mark size=4, mark options={solid,fill opacity=0}]
table {%
0 -26.1165657043457
5 -27.705530166626
10 -29.8844013214111
20 -32.2440185546875
30 -32.451358795166
};
\addlegendentry{$\text{PF}$}
\addplot [thick, blue, mark=o, mark size=4, mark options={solid,fill opacity=0}]
table {%
0 -22.4368457794189
5 -26.1798057556152
10 -27.9666118621826
20 -29.9004688262939
30 -30.0849533081055
};
\addlegendentry{$\text{pDANSE}  (\kappa=0\%)$}
\end{axis}

\end{tikzpicture}}
    \caption{NMSE (in dB) on $\Dataset_{\text{test}}$ vs. SMNR (in dB), demonstrating the performance of $\pdanse$ $(\kappa = 0\%)$ against the PF for the BSE task using noisy, high-dimensional measurements of the Lorenz-$63$ process. The nonlinear function is defined in \eqref{eq:highdim_meas_fn}. The PF is run with $100$ particles at the time of inference. $\pdanse$ was trained using a training dataset consisting of $N=1000$ trajectories of length $T=200$ each.}
    \label{fig:nmse_highdim_camera}
\end{figure}

\subsubsection{Camera model nonlinearity}\label{sec:results_highdim}

We now investigate the use of camera nonlinearity that provides high-dimensional measurements. Our inspiration for this investigation stems from the experimental setup described in \cite{buchnik2023latent, van2025deep}, where camera nonlinearity was used. The  $3$-dimensional state vector of the Lorenz-$63$ attractor is transformed using a Gaussian point spread function into a high-dimensional image, thereby mimicking a camera acquisition model. In our case, we transformed a $3$-dimensional $\mathbf{x}_t$ to an $8 \times 8$ image, or equivalently $64$-dimensional measurements. Note that one can also choose larger image sizes, e.g. $28 \times 28$. In this case, we opt for an image size of $8 \times 8$ for two reasons. The main reason is that it corresponds to a low-resolution image, representing a cheap and noisy camera model. That makes the estimation task challenging. The second reason is that processing high-resolution images requires a high computational resource, and hence, we avoid that. The $j^{\text{th}}$ noise-free measurement component is constructed as follows:
\begin{IEEEeqnarray}{rl}
h_{j}(\bstate_t; \pmb{\chi}^{(j)}) = 10 \exp \left( - \frac{1}{2\state_{t,3}} \bigg\Vert \begin{bmatrix}
\chi^{(j)}_{1} - \state_{t,1} \\
\chi^{(j)}_{2} - \state_{t,2} \\
\end{bmatrix}\bigg\Vert_{2}^{2}\right).
\label{eq:highdim_meas_fn}
\end{IEEEeqnarray}
Using \eqref{eq:highdim_meas_fn}, for $j=1, \ldots, 64$, one can generate a complete noise-free measurement vector $\mathbf{h}(\bstate_t) \in \mathbb{R}^{64}$. In \eqref{eq:highdim_meas_fn}, the $3$-dimensional state $\bstate_{t} \in \mathbb{R}^{3}$ is mapped using a $2$-dimensional reference point $\pmb{\chi}^{(j)} = \begin{bmatrix}
\chi^{(j)}_{1} & \chi^{(j)}_{2}
\end{bmatrix}^{\top}$ s.t. $\chi^{(j)}_{1} \in \left[-30, 30\right]$, $\chi^{(j)}_{2} \in \left[-40, 40\right]$. Both the ranges $\left[-30, 30\right]$ and $\left[-40, 40\right]$ are divided into $8$ equally spaced grid points. This creates a $2$-dimensional image grid of size $8 \times 8$. 
%The $2$-dimensional grid of size $8 \times 8$ is chosen by uniformly picking reference points $(\chi^{(j)}_{1}, \chi^{(j)}_{2})$ from the region $\left[-30, 30\right] \times \left[-40, 40\right]$ resulting in 
The measurement function $\bhn$ thus maps $\setR^{3} \to \setR^{64}$. Subsequently, we added Gaussian measurement noise $\bmnoise_{t} \in \setR^{64}$, s.t. $\bmnoise_{t} \sim \normaldist{\bmnoise_{t}}{\boldsymbol{0}}{\sigma^{2}_{\mnoise} \mathbf{I}_{64}}$ to create noisy measurements $\bmeas_{t} \in \setR^{64}$ following \eqref{eq:measurementsys}. Overall, a measurement sequence is a sequence of images, like a video. 

In Fig. \ref{fig:meas-highdim-reconstructed}, we provide a visual illustration of the performance of $\pdanse$ at $\text{SMNR}=0 \text{ dB}$. $\pdanse$ uses $L = 10$ MC samples during training and testing. Note that, in this case, we have measurements that are low-resolution images with a considerable strength of noise (signal and noise powers are the same), and still unsupervised learning provides a reasonable performance. We also did NMSE-versus-SMNR performance comparison with PF, {shown in Fig. \ref{fig:nmse_highdim_camera},} and that has a similar trend to Fig. \ref{fig:pdanse_cubic_3d}. We skip to show the NMSE-versus-SMNR plots due to brevity. Upon repeated experiments, we find that, in some instances, the unknown state can be estimated by $\pdanse$ only up to a constant phase factor.

\begin{figure}[t]
    \centering
    % This file was created with tikzplotlib v0.10.1.
\begin{tikzpicture}

\definecolor{darkgray176}{RGB}{176,176,176}
\definecolor{darkturquoise0191191}{RGB}{0,191,191}
\definecolor{darkviolet1910191}{RGB}{191,0,191}
\definecolor{green01270}{RGB}{0,127,0}
\definecolor{lightgray204}{RGB}{204,204,204}

\begin{axis}[
legend cell align={left},
legend style={fill opacity=0.8, draw opacity=1, text opacity=1, draw=lightgray204},
tick align=outside,
tick pos=left,
x grid style={darkgray176},
xlabel={SMNR (in dB)},
xmajorgrids,
xmin=-1.5, xmax=31.5,
xtick style={color=black},
xtick={-5,0,5,10,15,20,25,30,35},
xticklabels={
  \(\displaystyle {\ensuremath{-}5}\),
  \(\displaystyle {0}\),
  \(\displaystyle {5}\),
  \(\displaystyle {10}\),
  \(\displaystyle {15}\),
  \(\displaystyle {20}\),
  \(\displaystyle {25}\),
  \(\displaystyle {30}\),
  \(\displaystyle {35}\)
},
y grid style={darkgray176},
ylabel={NMSE (in dB)},
ymajorgrids,
ymin=-31.1478173077106, ymax=-0.827433294057846,
ytick style={color=black},
ytick={-35,-30,-25,-20,-15,-10,-5,0},
yticklabels={
  \(\displaystyle {\ensuremath{-}35}\),
  \(\displaystyle {\ensuremath{-}30}\),
  \(\displaystyle {\ensuremath{-}25}\),
  \(\displaystyle {\ensuremath{-}20}\),
  \(\displaystyle {\ensuremath{-}15}\),
  \(\displaystyle {\ensuremath{-}10}\),
  \(\displaystyle {\ensuremath{-}5}\),
  \(\displaystyle {0}\)
},
width=\linewidth,
height=0.75\linewidth
]
\path [draw=red, thick]
(axis cs:0,-13.3146899938583)
--(axis cs:0,-9.77804720401764);

\path [draw=red, thick]
(axis cs:10,-19.5290894508362)
--(axis cs:10,-12.7117323875427);

\path [draw=red, thick]
(axis cs:20,-26.621856212616)
--(axis cs:20,-16.4905247688293);

\path [draw=red, thick]
(axis cs:30,-29.7696180343628)
--(axis cs:30,-18.228253364563);

\path [draw=blue, thick]
(axis cs:0,-4.4349924325943)
--(axis cs:0,-2.2056325674057);

\path [draw=blue, thick]
(axis cs:10,-14.642865896225)
--(axis cs:10,-13.6469671726227);

\path [draw=blue, thick]
(axis cs:20,-15.9960757493973)
--(axis cs:20,-15.2885578870773);

\path [draw=blue, thick]
(axis cs:30,-13.2419065535069)
--(axis cs:30,-12.6969400346279);

\path [draw=green01270, thick]
(axis cs:0,-11.8830791711807)
--(axis cs:0,-10.4244064092636);

\path [draw=green01270, thick]
(axis cs:10,-18.960890352726)
--(axis cs:10,-17.5477537512779);

\path [draw=green01270, thick]
(axis cs:20,-22.2318145632744)
--(axis cs:20,-20.5013855099678);

\path [draw=green01270, thick]
(axis cs:30,-25.7863723039627)
--(axis cs:30,-24.5864762067795);

\path [draw=darkviolet1910191, thick]
(axis cs:0,-11.6380559802055)
--(axis cs:0,-10.2281149029732);

\path [draw=darkviolet1910191, thick]
(axis cs:10,-18.1471303701401)
--(axis cs:10,-16.8287912607193);

\path [draw=darkviolet1910191, thick]
(axis cs:20,-24.2895039916039)
--(axis cs:20,-22.8469782471657);

\path [draw=darkviolet1910191, thick]
(axis cs:30,-26.1747560501099)
--(axis cs:30,-24.9699277877808);

\path [draw=darkturquoise0191191, thick]
(axis cs:0,-11.0811555981636)
--(axis cs:0,-9.53590601682663);

\path [draw=darkturquoise0191191, thick]
(axis cs:10,-17.8826231360435)
--(axis cs:10,-16.2501893639565);

\path [draw=darkturquoise0191191, thick]
(axis cs:20,-24.2848395109177)
--(axis cs:20,-22.3964272737503);

\path [draw=darkturquoise0191191, thick]
(axis cs:30,-26.9892257452011)
--(axis cs:30,-25.5470367670059);

\addplot [thick, red, mark=diamond, mark size=5, mark options={solid,fill opacity=0}]
table {%
0 -11.546368598938
10 -16.1204109191895
20 -21.5561904907227
30 -23.9989356994629
};
\addlegendentry{PF}
\addplot [thick, blue, mark=o, mark size=4, mark options={solid,fill opacity=0}]
table {%
0 -3.3203125
10 -14.1449165344238
20 -15.6423168182373
30 -12.9694232940674
};
\addlegendentry{$\textrm{pDANSE}$ $(\kappa=0\%)$}
\addplot [thick, green01270, mark=asterisk, mark size=4, mark options={solid,fill opacity=0}]
table {%
0 -11.1537427902222
10 -18.254322052002
20 -21.3666000366211
30 -25.1864242553711
};
\addlegendentry{$\textrm{pDANSE}$ $(\kappa=1\%)$}
\addplot [thick, darkviolet1910191, mark=square, mark size=4, mark options={solid,fill opacity=0}]
table {%
0 -10.9330854415894
10 -17.4879608154297
20 -23.5682411193848
30 -25.5723419189453
};
\addlegendentry{$\textrm{pDANSE}$ $(\kappa=2\%)$}
\addplot [thick, darkturquoise0191191, mark=pentagon, mark size=4, mark options={solid,fill opacity=0}]
table {%
0 -10.3085308074951
10 -17.06640625
20 -23.340633392334
30 -26.2681312561035
};
\addlegendentry{$\textrm{pDANSE}$ $(\kappa=5\%)$}
\end{axis}

\end{tikzpicture}
    \caption{NMSE (in dB) on $\Dataset_{\text{test}}$ vs. SMNR (in dB), demonstrating the performance of $\pdanse$ ($\kappa \geq 0\%$) vis-\'a-vis the PF for the BSE task using noisy, half-wave rectified measurements of the Lorenz-$63$ process. The nonlinear function is defined in \eqref{eq:relu_measurement_fn}. While $\pdanse$ $(\kappa=0\%)$ underperforms, $\pdanse$ $(\kappa > 0\%)$ perform quite satisfactorily compared to the PF.}
    \label{fig:nmse_relu_pdanse}
\end{figure}
\subsection{Experiments in semi-supervised learning setups}\label{sec:resultssemisupervisedlearning}
In this section, we now show the BSE performance of $\pdanse$ for two different nonlinearities in the semi-supervised learning setup. Here we vary $\kappa$ and examine the performance trends. Here, the nonlinearities are: an element-wise, half-wave rectification nonlinearity and a Cartesian-to-spherical nonlinearity \cite{li2001survey}. For training and testing of $\pdanse$, we use $L = 10$ MC samples as previously. For the PF, we experimentally found that $100$ particles were sufficient to provide a reasonably good benchmark comparison. 

\subsubsection{Half-wave rectification nonlinearity}\label{sec:results_rectified}
Here, we present the BSE performance of the proposed $\pdanse$ method, using noisy, half-wave rectified measurements. %We remark that similar experiments have also been shown in our prior work \cite[Section IV.]{ghosh2025pdanse}. 
Following \eqref{eq:measurementsys}, the elements of the measurement function $\bhn(\bstate_t)$ in this case are  defined as follows:
\begin{IEEEeqnarray}{rl}\label{eq:relu_measurement_fn}
    h_{i}\left(\bstate_t\right) = \max\left(0, \state_{t,i}\right), \, \, \, i=1, 2, 3.
\label{eq:lorenzssm_hfn}
\end{IEEEeqnarray}
%\eqref{eq:relu_measurement_fn} implies that the information about the negative components of the state is lost in the noisy measurement.

We show the quantitative BSE performance of $\pdanse$ at different $\kappa$ values versus the model-driven PF in Fig. \ref{fig:nmse_relu_pdanse}. Similar experiments have also been shown in our prior work \cite[Section IV.]{ghosh2025pdanse}. We observe that the unsupervised version of $\pdanse$ $(\kappa = 0\%)$ performs poorly compared to the model-driven PF on the BSE task. Our ansatz is that this occurs due to the loss of information about the unknown state caused by the rectification nonlinearity \eqref{eq:relu_measurement_fn}, rendering unsupervised learning challenging. We overcome this challenge by using a limited amount of labelled data $(\kappa > 0 \%)$, and we observe that the semi-supervised versions of $\pdanse$ $(\kappa > 0 \%)$ perform quite close to that of the PF. Furthermore, we observe in Fig. \ref{fig:nmse_relu_pdanse} that there is significant improvement with just $\kappa = 1\%$, and there is no drastic improvement by increasing $\kappa=5\%$. This corroborates the efficiency of semi-supervised learning, where the usage of a limited amount of labeled data is helpful. 

%\begin{color}{blue}
\begin{figure}[t]
    \centering
    \scalebox{1.0}{% This file was created with tikzplotlib v0.10.1.
\begin{tikzpicture}

\definecolor{darkgray176}{RGB}{176,176,176}
\definecolor{lightgray204}{RGB}{204,204,204}
\definecolor{darkgray176}{RGB}{176,176,176}
\definecolor{darkturquoise0191191}{RGB}{0,191,191}
\definecolor{darkviolet1910191}{RGB}{191,0,191}
\definecolor{green01270}{RGB}{0,127,0}
\definecolor{lightgray204}{RGB}{204,204,204}

\begin{axis}[
legend cell align={left},
legend style={fill opacity=0.8, draw opacity=1, text opacity=1, draw=lightgray204},
tick align=outside,
tick pos=left,
x grid style={darkgray176},
xlabel={SMNR (in dB)},
xmajorgrids,
xmin=-1.5, xmax=31.5,
xtick style={color=black},
xtick={-5,0,5,10,15,20,25,30,35},
xticklabels={
  \(\displaystyle {\ensuremath{-}5}\),
  \(\displaystyle {0}\),
  \(\displaystyle {5}\),
  \(\displaystyle {10}\),
  \(\displaystyle {15}\),
  \(\displaystyle {20}\),
  \(\displaystyle {25}\),
  \(\displaystyle {30}\),
  \(\displaystyle {35}\)
},
y grid style={darkgray176},
ylabel={NMSE (in dB)},
ymajorgrids,
ymin=-31.0591481268406, ymax=-2.68948609232902,
ytick style={color=black},
ytick={-35,-30,-25,-20,-15,-10,-5,0},
yticklabels={
  \(\displaystyle {\ensuremath{-}35}\),
  \(\displaystyle {\ensuremath{-}30}\),
  \(\displaystyle {\ensuremath{-}25}\),
  \(\displaystyle {\ensuremath{-}20}\),
  \(\displaystyle {\ensuremath{-}15}\),
  \(\displaystyle {\ensuremath{-}10}\),
  \(\displaystyle {\ensuremath{-}5}\),
  \(\displaystyle {0}\)
},
width=0.95\linewidth,
height=0.75\linewidth
]
\path [draw=red, semithick]
(axis cs:0,-13.3146899938583)
--(axis cs:0,-9.77804720401764);

\path [draw=red, semithick]
(axis cs:10,-19.5290894508362)
--(axis cs:10,-12.7117323875427);

\path [draw=red, semithick]
(axis cs:20,-26.621856212616)
--(axis cs:20,-16.4905247688293);

\path [draw=red, semithick]
(axis cs:30,-29.7696180343628)
--(axis cs:30,-18.228253364563);

\path [draw=blue, semithick]
(axis cs:0,-11.6380705237389)
--(axis cs:0,-10.2277761101723);

\path [draw=blue, semithick]
(axis cs:10,-18.1460189819336)
--(axis cs:10,-16.8268241882324);

\path [draw=blue, semithick]
(axis cs:20,-24.2893719673157)
--(axis cs:20,-22.8423495292664);

\path [draw=blue, semithick]
(axis cs:30,-26.1668190956116)
--(axis cs:30,-24.9566550254822);

\path [draw=blue, semithick]
(axis cs:0,-4.98450767993927)
--(axis cs:0,-3.97901618480682);

\path [draw=blue, semithick]
(axis cs:10,-11.690207362175)
--(axis cs:10,-9.66957581043243);

\path [draw=blue, semithick]
(axis cs:20,-18.9448436498642)
--(axis cs:20,-15.8573795557022);

\path [draw=blue, semithick]
(axis cs:30,-23.3823227882385)
--(axis cs:30,-19.7754454612732);

\addplot [thick, red, mark=diamond, mark size=5, mark options={solid,fill opacity=0}]
table {%
0 -11.546368598938
10 -16.1204109191895
20 -21.5561904907227
30 -23.9989356994629
};
\addlegendentry{$\textrm{PF}$}
\addplot [thick, darkviolet1910191, mark=square, mark size=4, mark options={solid,fill opacity=0}]
table {%
0 -10.9329233169556
10 -17.486421585083
20 -23.565860748291
30 -25.5617370605469
};
\addlegendentry{$\textrm{pDANSE}$ $(\kappa=2\%)$}
\addplot [thick, dashed, darkviolet1910191, mark=square, mark size=4, mark options={solid,fill opacity=0}]
table {%
0 -4.48176193237305
10 -10.6798915863037
20 -17.4011116027832
30 -21.5788841247559
};
\addlegendentry{$\textrm{pDANSE-Sup.}$ $(\kappa=2\%)$}
\end{axis}

\end{tikzpicture}}
    \caption{{NMSE (in dB) on $\Dataset_{\text{test}}$ vs. SMNR (in dB), demonstrating the performance of pDANSE ($\kappa = 2\%$), pDANSE-Supervised ($\kappa = 2\%$) vis-\'a-vis the PF for the BSE task using noisy, half-wave rectified measurements of the Lorenz-$63$ process. The nonlinear function is defined in \eqref{eq:relu_measurement_fn}. While pDANSE-Supervised $(\kappa=2\%)$ underperforms, pDANSE $(\kappa=2\%)$ perform quite satisfactorily compared to the PF.}}
    \label{fig:nmse_relu_pdanse_w_supervsied}
\end{figure}
Experimentally, we use the semi-supervised learning setup described in Section \ref{sec:results_rectified} with noisy, half-wave rectified nonlinear measurements. Both pDANSE and pDANSE-Supervised were trained on $\Dataset_{\text{semi}}$ with $\kappa=2\%$, except that pDANSE-Supervised only used the supervised part of $\Dataset_{\text{semi}}$ corresponding to $\kappa=2\%$. The results are shown in Fig. \ref{fig:nmse_relu_pdanse_w_supervsied}. We find that pDANSE outperforms pDANSE-Supervised at $\kappa=2\%$, demonstrating the benefits of semi-supervised learning. 

Lastly, we performed a sensitivity analysis and tested $L= 1, 10, 50$ in the case of pDANSE. We used the experimental setup with semi-supervised learning together with a noisy, rectified measurement function described in Section \ref{sec:results_rectified}. The results are shown in Fig. \ref{fig:relu_pdanse_supervised_diff_nmc_relu}. We observe that the pDANSE with $\kappa=2\%$ with $L=10$ performs better than $L=1, 50$, and hence we use $L=10$ consistently throughout our experiments.  
\begin{figure}
    \centering
    \scalebox{1.0}{% This file was created with tikzplotlib v0.10.1.
\begin{tikzpicture}

\definecolor{darkgray176}{RGB}{176,176,176}
\definecolor{darkviolet1910191}{RGB}{191,0,191}
\definecolor{green01270}{RGB}{0,127,0}
\definecolor{lightgray204}{RGB}{204,204,204}

\begin{axis}[
legend cell align={left},
legend style={fill opacity=0.8, draw opacity=1, text opacity=1, draw=lightgray204},
tick align=outside,
tick pos=left,
x grid style={darkgray176},
xlabel={SMNR (in dB)},
xmajorgrids,
xmin=-1.5, xmax=31.5,
xtick style={color=black},
xtick={-5,0,5,10,15,20,25,30,35},
xticklabels={
  \(\displaystyle {\ensuremath{-}5}\),
  \(\displaystyle {0}\),
  \(\displaystyle {5}\),
  \(\displaystyle {10}\),
  \(\displaystyle {15}\),
  \(\displaystyle {20}\),
  \(\displaystyle {25}\),
  \(\displaystyle {30}\),
  \(\displaystyle {35}\)
},
y grid style={darkgray176},
ylabel={NMSE (in dB)},
ymajorgrids,
ymin=-27.4842610746622, ymax=-4.42090298831463,
ytick style={color=black},
ytick={-30,-25,-20,-15,-10,-5,0},
yticklabels={
  \(\displaystyle {\ensuremath{-}30}\),
  \(\displaystyle {\ensuremath{-}25}\),
  \(\displaystyle {\ensuremath{-}20}\),
  \(\displaystyle {\ensuremath{-}15}\),
  \(\displaystyle {\ensuremath{-}10}\),
  \(\displaystyle {\ensuremath{-}5}\),
  \(\displaystyle {0}\)
},
width=0.95\linewidth,
height=0.75\linewidth
]
\path [draw=red, semithick]
(axis cs:0,-14.5403207540512)
--(axis cs:0,-10.596329331398);

\path [draw=red, semithick]
(axis cs:10,-20.5690952837467)
--(axis cs:10,-20.0303843915462);

\path [draw=red, semithick]
(axis cs:20,-25.8402678966522)
--(axis cs:20,-18.4602158069611);

\path [draw=red, semithick]
(axis cs:30,-25.4776158332825)
--(axis cs:30,-19.0991778373718);

\path [draw=blue, semithick]
(axis cs:0,-6.97037589550018)
--(axis cs:0,-5.46923744678497);

\path [draw=blue, semithick]
(axis cs:10,-16.9720823764801)
--(axis cs:10,-14.6171624660492);

\path [draw=blue, semithick]
(axis cs:20,-24.0498322248459)
--(axis cs:20,-22.6911956071854);

\path [draw=blue, semithick]
(axis cs:30,-26.4359266161919)
--(axis cs:30,-24.8929994702339);

\path [draw=darkviolet1910191, semithick]
(axis cs:0,-11.6379261016846)
--(axis cs:0,-10.2275905609131);

\path [draw=darkviolet1910191, semithick]
(axis cs:10,-18.1465187072754)
--(axis cs:10,-16.829273223877);

\path [draw=darkviolet1910191, semithick]
(axis cs:20,-24.2907084822655)
--(axis cs:20,-22.8516712784767);

\path [draw=darkviolet1910191, semithick]
(axis cs:30,-26.1623842120171)
--(axis cs:30,-24.9608228802681);

\path [draw=green01270, semithick]
(axis cs:0,-8.63352936506271)
--(axis cs:0,-7.07477980852127);

\path [draw=green01270, semithick]
(axis cs:10,-16.6264808177948)
--(axis cs:10,-13.7314369678497);

\path [draw=green01270, semithick]
(axis cs:20,-24.1399022936821)
--(axis cs:20,-22.8137529492378);

\path [draw=green01270, semithick]
(axis cs:30,-26.4177566766739)
--(axis cs:30,-25.0188163518906);

\addplot [thick, red, mark=diamond, mark size=5, mark options={solid,fill opacity=0}]
table {%
0 -12.5683250427246
10 -20.2997398376465
20 -22.1502418518066
30 -22.2883968353271
};
\addlegendentry{PF}
\addplot [thick, blue, mark=o, mark size=5, mark options={solid,fill opacity=0}]
table {%
0 -6.21980667114258
10 -15.7946224212646
20 -23.3705139160156
30 -25.6644630432129
};
\addlegendentry{$\textrm{pDANSE}$ $(L=1)$}
\addplot [thick, darkviolet1910191, mark=square, mark size=5, mark options={solid,fill opacity=0}]
table {%
0 -10.9327583312988
10 -17.4878959655762
20 -23.5711898803711
30 -25.5616035461426
};
\addlegendentry{$\textrm{pDANSE}$ $(L=10)$}
\addplot [thick, green01270, mark=asterisk, mark size=5, mark options={solid,fill opacity=0}]
table {%
0 -7.85415458679199
10 -15.1789588928223
20 -23.47682762146
30 -25.7182865142822
};
\addlegendentry{$\textrm{pDANSE}$ $(L=50)$}
\end{axis}

\end{tikzpicture}}
    \caption{{NMSE (in dB) on $\Dataset_{\text{test}}$ vs. SMNR (in dB), demonstrating the performance of pDANSE ($\kappa = 2\%$) for different $L=1, 10, 50$, vis-\'a-vis the PF for the BSE task using noisy, half-wave rectified measurements of the Lorenz-$63$ process. The nonlinear function is defined in \eqref{eq:relu_measurement_fn}.}}
\label{fig:relu_pdanse_supervised_diff_nmc_relu}
\end{figure}

\begin{figure*}[t]
    \centering
    \includegraphics[width=\textwidth]{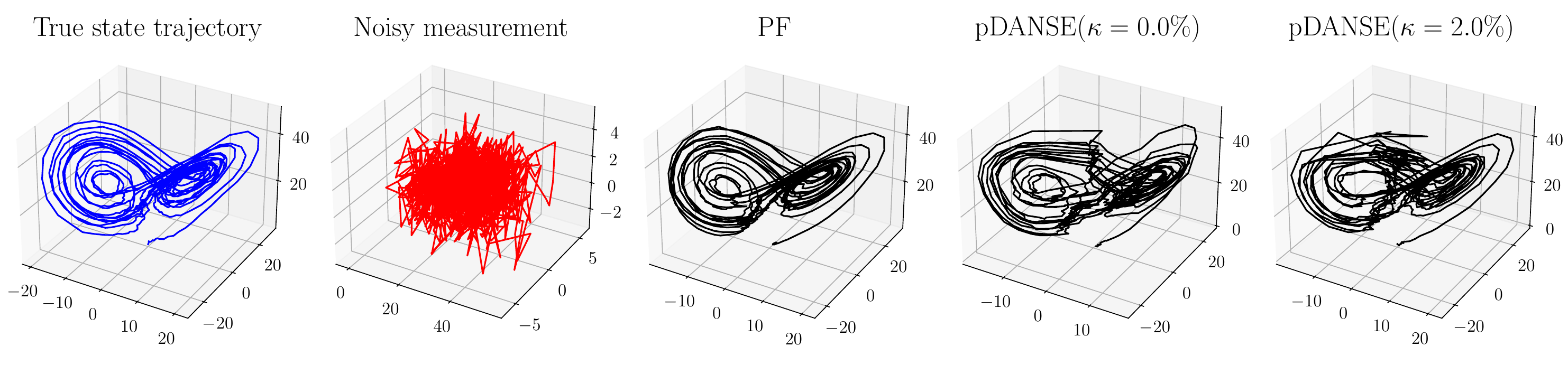}
    \caption{{Visual illustration of the performance of pDANSE for different $\kappa$ for the BSE problem using noisy, Cartesian-to-spherical  measurements of the Lorenz-$63$ process, at $\text{SMNR} = 20 \text{ dB}$. A shorter trajectory length $T_{\text{test}} = 1000$ is shown for clarity and ease of understanding. Far left - An example of a true state trajectory is shown in blue. Middle left - The corresponding noisy measurement (shown in red). Middle - The estimated state trajectory (shown in black) is the posterior mean estimates using the PF. Middle right and far right - The estimated state trajectories (shown in black) are the corresponding posterior mean estimates using PF and pDANSE at $\kappa = 0\%$ and $\kappa = 2\%$. }
    \label{fig:pdanse_cart2sph3d_3d_1col_}}
\end{figure*}
\begin{comment}
\begin{figure}[t]
    \centering
    \input{figs/journalpaper/cart2sph3d/nmse_pf_pdanse_nsup_all_cart2sph3dmod}
    \caption{NMSE (in dB) on $\Dataset_{\text{test}}$ vs. SMNR (in dB), demonstrating the performance of $\pdanse$ ($\kappa= 0\%)$ and $\pdanse$ ($\kappa= 2\%)$ for the BSE task using noisy, Cartesian-to-spherical measurements of the Lorenz-$63$ process. The nonlinear function is defined in \eqref{eq:cart2sph3d_meas_fn}. The comparison is against the model-driven PF.}
    \label{fig:nmse_cart2sph3d_pdanse}
\end{figure}
\end{comment}

\begin{figure}[t]
    \centering
    % This file was created with tikzplotlib v0.10.1.
\begin{tikzpicture}

\definecolor{darkgray176}{RGB}{176,176,176}
\definecolor{darkviolet1910191}{RGB}{191,0,191}
\definecolor{lightgray204}{RGB}{204,204,204}

\begin{axis}[
legend cell align={left},
legend style={fill opacity=0.8, draw opacity=1, text opacity=1, draw=lightgray204},
tick align=outside,
tick pos=left,
x grid style={darkgray176},
xlabel={SMNR (in dB)},
xmajorgrids,
xmin=-1.5, xmax=31.5,
xtick style={color=black},
xtick={-5,0,5,10,15,20,25,30,35},
xticklabels={
  \(\displaystyle {\ensuremath{-}5}\),
  \(\displaystyle {0}\),
  \(\displaystyle {5}\),
  \(\displaystyle {10}\),
  \(\displaystyle {15}\),
  \(\displaystyle {20}\),
  \(\displaystyle {25}\),
  \(\displaystyle {30}\),
  \(\displaystyle {35}\)
},
y grid style={darkgray176},
ylabel={NMSE (in dB)},
ymajorgrids,
ymin=-31.8889796495438, ymax=-0.632314467430115,
ytick style={color=black},
ytick={-35,-30,-25,-20,-15,-10,-5,0},
yticklabels={
  \(\displaystyle {\ensuremath{-}35}\),
  \(\displaystyle {\ensuremath{-}30}\),
  \(\displaystyle {\ensuremath{-}25}\),
  \(\displaystyle {\ensuremath{-}20}\),
  \(\displaystyle {\ensuremath{-}15}\),
  \(\displaystyle {\ensuremath{-}10}\),
  \(\displaystyle {\ensuremath{-}5}\),
  \(\displaystyle {0}\)
},
width=\linewidth,
height=0.75\linewidth
]
\path [draw=red, semithick]
(axis cs:0,-8.45058465003967)
--(axis cs:0,-6.33879446983337);

\path [draw=red, semithick]
(axis cs:10,-16.7247097492218)
--(axis cs:10,-9.80679774284363);

\path [draw=red, semithick]
(axis cs:20,-24.860981464386)
--(axis cs:20,-12.3047366142273);

\path [draw=red, semithick]
(axis cs:30,-30.4682221412659)
--(axis cs:30,-16.9258208274841);

\path [draw=blue, semithick]
(axis cs:0,-2.72163581848145)
--(axis cs:0,-2.05307197570801);

\path [draw=blue, semithick]
(axis cs:10,-13.096010684967)
--(axis cs:10,-10.5998358726501);

\path [draw=blue, semithick]
(axis cs:20,-20.6504509449005)
--(axis cs:20,-20.1247565746307);

\path [draw=blue, semithick]
(axis cs:30,-23.3174862414598)
--(axis cs:30,-22.8264804333448);

\path [draw=darkviolet1910191, semithick]
(axis cs:0,-5.08455844223499)
--(axis cs:0,-4.84060101211071);

\path [draw=darkviolet1910191, semithick]
(axis cs:10,-12.8024022579193)
--(axis cs:10,-11.2013857364655);

\path [draw=darkviolet1910191, semithick]
(axis cs:20,-17.4801238626242)
--(axis cs:20,-17.0029236227274);

\path [draw=darkviolet1910191, semithick]
(axis cs:30,-19.0278548002243)
--(axis cs:30,-18.3880311250687);

\addplot [thick, red, mark=diamond, mark size=5, mark options={solid,fill opacity=0}]
table {%
0 -7.39468955993652
10 -13.2657537460327
20 -18.5828590393066
30 -23.697021484375
};
\addlegendentry{$\textrm{PF}$ }
\addplot [thick, blue, mark=o, mark size=5, mark options={solid,fill opacity=0}]
table {%
0 -2.38735389709473
10 -11.8479232788086
20 -20.3876037597656
30 -23.0719833374023
};
\addlegendentry{$\textrm{pDANSE}$ $(\kappa=0\%)$}
\addplot [thick, darkviolet1910191, mark=square, mark size=5, mark options={solid,fill opacity=0}]
table {%
0 -4.96257972717285
10 -12.0018939971924
20 -17.2415237426758
30 -18.7079429626465
};
\addlegendentry{$\textrm{pDANSE}$ $(\kappa=2\%)$}
\end{axis}

\end{tikzpicture}
    \caption{{NMSE (in dB) on $\Dataset_{\text{test}}$ vs. SMNR (in dB), demonstrating the performance of pDANSE ($\kappa= 0\%)$ and pDANSE ($\kappa= 2\%)$ for the BSE task using noisy, Cartesian-to-spherical measurements of the Lorenz-$63$ process. The nonlinear function is defined in \eqref{eq:cart2sph3d_meas_fn}. The comparison is against the model-driven PF.}}
    \label{fig:nmse_cart2sph3d_pdanse}
\end{figure}

\begin{comment}
\begin{figure}[t]
    \centering
    \input{figs/journalpaper/cart2sph3d/LoreznSSM_Axiswise_pdanse_nsup_pf_smnr10.0dB_plot}
    \caption{Demonstrating the performance of $\pdanse$ for different $\kappa$ for the BSE problem using noisy, Cartesian-to-spherical  measurements of the Lorenz-$63$ process, at $\text{SMNR} = 10 \text{ dB}$ and $\sigmapnoise^2$ corresponding to $-10$ dB. The estimated state trajectories (shown in black) are the corresponding posterior mean estimates using PF and $\pdanse$ at different $\kappa$ values. The PF is model-driven and knows the exact state dynamics.}
    \label{fig:pdanse_cart2sph3d_3d}
\end{figure}
\end{comment}
\subsubsection{Cartesian-to-spherical nonlinearity}\label{sec:results_cart2sph3d}
Next, we consider the Cartesian-to-spherical coordinate transformation function as a nonlinearity. Following \eqref{eq:measurementsys}, the elements of the measurement function $\bhn(\bstate_t)$ in this case are defined as follows: 
\begin{IEEEeqnarray}{rl}
    \mathbf{h}(\bstate_t) = \begin{bmatrix} h_{1}(\bstate_t) \\ h_{2}(\bstate_t) \\ h_{3}(\bstate_t) \end{bmatrix} = \begin{bmatrix} \sqrt{\sum_{i=1}^{3} \state_{t,i}^2}\\ \text{tan}^{-1}\left(\frac{\state_{t,2}}{\state_{t,1}}\right) \\ \text{tan}^{-1}\left(\frac{\state_{t,3}}{\sqrt{\state_{t,1}^2 + \state_{t,2}^2}}\right) \\ \end{bmatrix}.
\label{eq:cart2sph3d_meas_fn}
\end{IEEEeqnarray}
The measurement function in \eqref{eq:cart2sph3d_meas_fn} is of significance in target tracking, as in the case of a dish radar \cite{li2001survey}. 

This form of the nonlinearity in \eqref{eq:cart2sph3d_meas_fn} renders the BSE task quite challenging. The nonlinearity has phase wrapping components in the second and third elements in \eqref{eq:cart2sph3d_meas_fn}. {We visually illustrate the BSE performance of PF and $\pdanse$ in Fig. \ref{fig:pdanse_cart2sph3d_3d_1col_} at $\text{SMNR}=20$ dB. The BSE task appears quite challenging for both the semi-supervised and unsupervised versions, since they are uninformed about the STM and the phase wrapping components. We observe that PF, which knows the underlying STM, performs quite well. Also, $\pdanse$ $(\kappa = 0\%)$ and $\pdanse$ $(\kappa = 2\%)$ perform reasonably well compared to the PF. }

Next, we show NMSE versus SMNR plots for $\pdanse$ against PF in Fig. \ref{fig:nmse_cart2sph3d_pdanse}. Both the unsupervised $\pdanse$ ($\kappa = 0\%$) and the semi-supervised $\pdanse$ ($\kappa=2\%$) are compared against the model-driven PF. $\pdanse$ provides similar performance to that of the PF.
\subsection{Experiments with a high-dimensional Lorenz-$96$ system}\label{sec:results_highdim_lorenz96_relu}
We conducted further experiments on a stochastic, $20$-dimensional Lorenz-$96$ system with nonlinear half-wave rectified measurements. This system has been used before for forecasting tasks in \cite{lin2025efficient}. The measurement system follows the same as \eqref{eq:measurementsys}, with the measurement function 
\begin{equation}\label{eq:relu_measurement_fn_lorenz96}
    h_{i}\left(\bstate_t\right) = \max\left(0, \state_{t,i}\right), \, \, \, i=1, 2, \ldots, 20.
\end{equation}
For our experiments, we generate noisy test data where we use $\bmnoiseCov = \sigma^2_{\mnoise} \mathbf{I}_{20}$; here $\sigma^2_{\mnoise}$ is chosen using the average signal-to-measurement noise ratio (SMNR) in dB scale  \cite[section III-C]{ghosh2023dansejrnl}. 
The Lorenz-$96$ system details can be found in \cite[Appendix B]{ghosh2023dansejrnl}, and we followed a similar experimental setup as described in \cite[Section H]{ghosh2023dansejrnl}. 
\begin{figure}[t]
    \centering
    \scalebox{1.0}{% This file was created with tikzplotlib v0.10.1.
\begin{tikzpicture}

\definecolor{darkgray176}{RGB}{176,176,176}
\definecolor{darkviolet1910191}{RGB}{191,0,191}
\definecolor{lightgray204}{RGB}{204,204,204}

\begin{axis}[
legend cell align={left},
legend style={fill opacity=0.8, draw opacity=1, text opacity=1, draw=lightgray204},
tick align=outside,
tick pos=left,
x grid style={darkgray176},
xlabel={SMNR (in dB)},
xmajorgrids,
xmin=-1.5, xmax=31.5,
xtick style={color=black},
xtick={-5,0,5,10,15,20,25,30,35},
xticklabels={
  \(\displaystyle {\ensuremath{-}5}\),
  \(\displaystyle {0}\),
  \(\displaystyle {5}\),
  \(\displaystyle {10}\),
  \(\displaystyle {15}\),
  \(\displaystyle {20}\),
  \(\displaystyle {25}\),
  \(\displaystyle {30}\),
  \(\displaystyle {35}\)
},
y grid style={darkgray176},
ylabel={NMSE (in dB)},
ymajorgrids,
ymin=-18.2055130340159, ymax=-2.91494924947619,
ytick style={color=black},
ytick={-20,-18,-16,-14,-12,-10,-8,-6,-4,-2},
yticklabels={
  \(\displaystyle {\ensuremath{-}20}\),
  \(\displaystyle {\ensuremath{-}18}\),
  \(\displaystyle {\ensuremath{-}16}\),
  \(\displaystyle {\ensuremath{-}14}\),
  \(\displaystyle {\ensuremath{-}12}\),
  \(\displaystyle {\ensuremath{-}10}\),
  \(\displaystyle {\ensuremath{-}8}\),
  \(\displaystyle {\ensuremath{-}6}\),
  \(\displaystyle {\ensuremath{-}4}\),
  \(\displaystyle {\ensuremath{-}2}\)
},
width=\linewidth,
height=0.75\linewidth
]
\path [draw=red, semithick]
(axis cs:0,-3.93010233342648)
--(axis cs:0,-3.60997487604618);

\path [draw=red, semithick]
(axis cs:10,-7.23932099342346)
--(axis cs:10,-6.36657881736755);

\path [draw=red, semithick]
(axis cs:20,-10.0401799082756)
--(axis cs:20,-9.01942282915115);

\path [draw=red, semithick]
(axis cs:30,-11.4073729515076)
--(axis cs:30,-10.3274998664856);

\path [draw=blue, semithick]
(axis cs:0,-9.61459168791771)
--(axis cs:0,-9.01088133454323);

\path [draw=blue, semithick]
(axis cs:10,-15.6204179823399)
--(axis cs:10,-15.0567212998867);

\path [draw=blue, semithick]
(axis cs:20,-12.4432728290558)
--(axis cs:20,-12.0081317424774);

\path [draw=blue, semithick]
(axis cs:30,-10.9767584502697)
--(axis cs:30,-10.5527768433094);

\path [draw=darkviolet1910191, semithick]
(axis cs:0,-9.45501390099525)
--(axis cs:0,-8.84474310278893);

\path [draw=darkviolet1910191, semithick]
(axis cs:10,-17.5104874074459)
--(axis cs:10,-16.6637755930424);

\path [draw=darkviolet1910191, semithick]
(axis cs:20,-16.138001292944)
--(axis cs:20,-15.2394782602787);

\path [draw=darkviolet1910191, semithick]
(axis cs:30,-11.0771287828684)
--(axis cs:30,-10.5841833204031);

\addplot [thick, red, mark=diamond, mark size=5, mark options={solid,fill opacity=0}]
table {%
0 -3.77003860473633
10 -6.80294990539551
20 -9.52980136871338
30 -10.8674364089966
};
\addlegendentry{$\textrm{EnKF}$ }
\addplot [thick, blue, mark=o, mark size=5, mark options={solid,fill opacity=0}]
table {%
0 -9.31273651123047
10 -15.3385696411133
20 -12.2257022857666
30 -10.7647676467896
};
\addlegendentry{$\textrm{pDANSE}$ $(\kappa=0\%)$}
\addplot [thick, darkviolet1910191, mark=square, mark size=5, mark options={solid,fill opacity=0}]
table {%
0 -9.14987850189209
10 -17.0871315002441
20 -15.6887397766113
30 -10.8306560516357
};
\addlegendentry{$\textrm{pDANSE}$ $(\kappa=8\%)$}
\end{axis}

\end{tikzpicture}}
    \caption{{The average NMSE (in dB) on $\Dataset_{\text{test}}$ versus $\text{SMNR}$ performances to illustrate the success of pDANSE ($\kappa=8\%$) for the $20$-dimensional Lorenz-$96$ system with noisy, half-wave rectified measurements. The comparison is against the unsupervised pDANSE ($\kappa=0\%$) and a model-driven ensemble Kalman filter (EnKF).}}
    \label{fig:nmse_vs_smnr_relu_lorenz96}
\end{figure}
\begin{figure}[t]
    \centering
    \includegraphics[width=\linewidth]{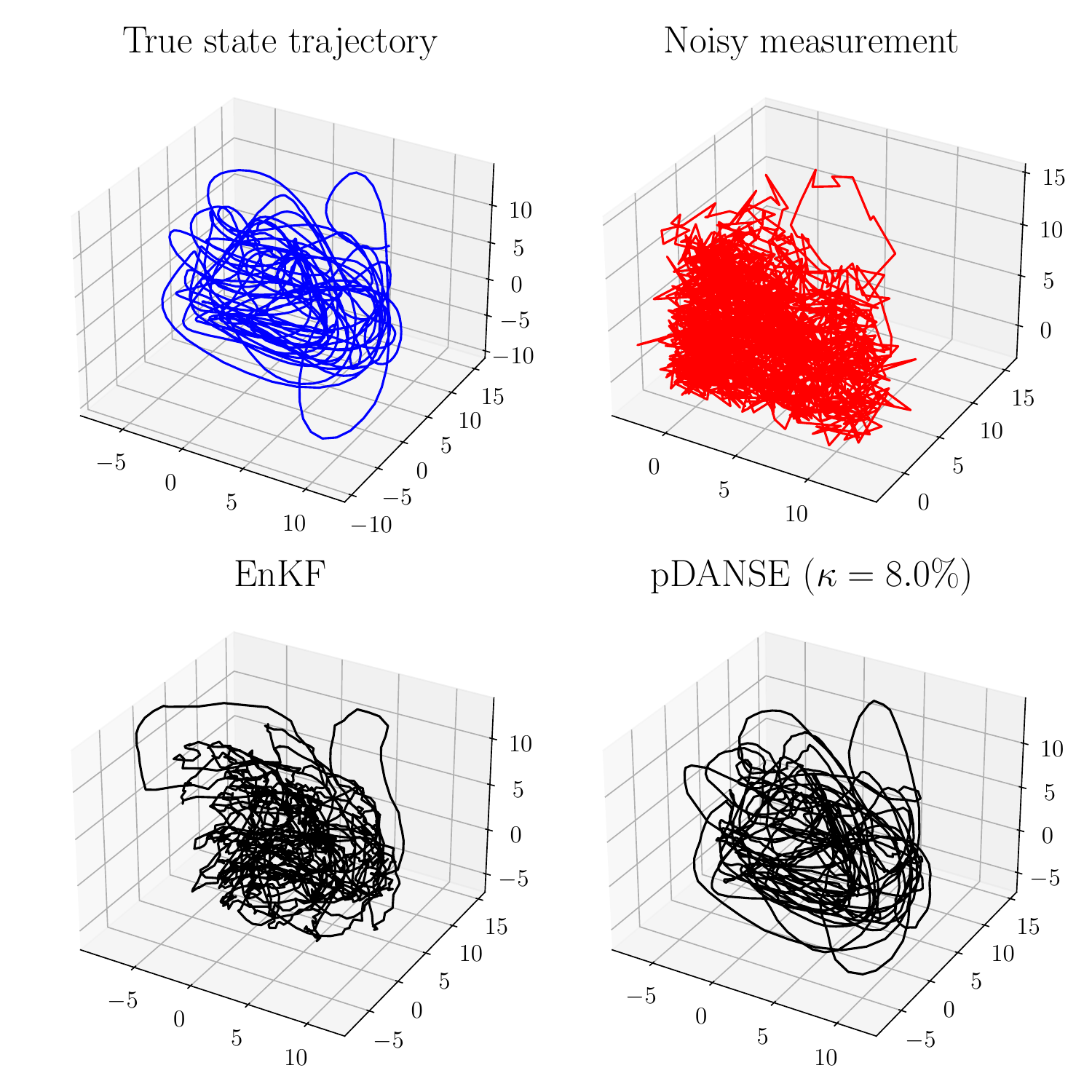}
    \caption{{Visual illustration of the BSE performance for $20$-dimensional Lorenz-$96$ systems with half-wave, rectified measurements in a semi-supervised learning setup at $\text{SMNR} = 10 \text{ dB}$. For the ease of illustration, only the first $3$ out $20$ state dimensions are shown in the figure. Top-left - A true state trajectory for the stochastic Lorenz-$63$ system from $\Dataset_{\text{test}}$. Top-right - The corresponding noisy measurement trajectory. Bottom-left - The estimated trajectory of EnKF. Bottom-right- The estimated trajectory of pDANSE. %Demonstrating the performance of $\pdanse$ using $\kappa = 0.0$ for the BSE problem using noisy, element-wise cubic measurements of the Lorenz-$63$ process, at $\text{SMNR} = 20 \text{ dB}$. The estimated state trajectories (shown in black) are the corresponding posterior mean estimates using PF and $\pdanse$ $(\kappa=0\%)$. The PF is model-driven and knows the exact state dynamics.
    }}
    \label{fig:pdanse_relu_lorenz96_3d}
\end{figure}
The results are shown in Fig. \ref{fig:nmse_vs_smnr_relu_lorenz96}. We also implemented a model-driven ensemble Kalman filter (EnKF) with covariance inflation and localization as in \cite{roth2017ensemble, chen2022autodifferentiable} for comparison. We used $100$ ensembles for the EnKF and experimentally found it useful to have adaptively higher ensemble values at higher SMNR values. We observe that the EnKF performs quite poorly at lower SMNR values compared to pDANSE (at $\kappa=0\%, 8\%$). We surmise that the reason for this poor performance is possibly due to a suboptimal manner of our implementation of the localization radius and covariance inflation (we experimentally observed that the EnKF is quite difficult to numerically implement at lower SMNR values). Despite this, it is clear that the proposed pDANSE can also work with high-dimensional dynamical systems. Also, as we observed for Lorenz-$63$, in the case of rectified measurements, there are performance improvements with the semi-supervised learning setup compared with the unsupervised one. We also provide a qualitative comparison of this result in Fig. \ref{fig:pdanse_relu_lorenz96_3d}, where we observe similar trends as in Fig. \ref{fig:nmse_vs_smnr_relu_lorenz96}, where pDANSE ($\kappa=2\%$) outperforms EnKF in the low SMNR regime.  
%\end{color}
\section{Conclusion}\label{sec:conclusion}
This work presents particle DANSE ($\pdanse$) -- a method that utilizes an RNN-based Gaussian prior, which, together with a known, nonlinear measurement system, can handle the task of Bayesian state estimation for a nonlinear dynamical process without requiring knowledge of the state transition model. We show that $\pdanse$ can be trained in an unsupervised manner and, if necessary, also in a semi-supervised manner. Owing to the presence of a nonlinearity in the measurement system, $\pdanse$ can only provide second-order statistics of the posterior distribution at the time of inference. Using the benchmark stochastic Lorenz-$63$ system, {we compare $\pdanse$ against the model-driven PF}. Using four nonlinear measurement systems, we show that $\pdanse$ offers competitive Bayesian state estimation performance compared to that of the PF. {Additionally, we also show that pDANSE works for high-dimensional stochastic Lorenz-$96$ system and can outperform EnKF in the low SMNR regime. }

\bibliographystyle{IEEEbib}
\bibliography{main}

\begin{thebibliography}{10}

\bibitem{dehganpour2019survey}
K.~Dehghanpour, Z.~Wang, J.~Wang, Y.~Yuan, and F.~Bu,
\newblock ``A survey on state estimation techniques and challenges in smart
  distribution systems,''
\newblock {\em IEEE Transactions on Smart Grid}, vol. 10, no. 2, pp.
  2312--2322, 2019.

\bibitem{shlezinger2025survey}
N.~Shlezinger, G.~Revach, A.~Ghosh, S.~Chatterjee, S.~Tang, T.~Imbiriba,
  J.~Dunik, O.~Straka, P.~Closas, and Y.~C. Eldar,
\newblock ``{Artificial Intelligence-Aided Kalman Filters: AI-Augmented Designs
  for Kalman-Type Algorithms},''
\newblock {\em IEEE Signal Processing Magazine}, pp. 2--26, 2025.

\bibitem{gustafsson2010particle}
Fredrik Gustafsson,
\newblock ``Particle filter theory and practice with positioning
  applications,''
\newblock {\em IEEE Aerospace and Electronic Systems Magazine}, vol. 25, no. 7,
  pp. 53--82, 2010.

\bibitem{ghosh2023dansejrnl}
A.~Ghosh, A.~Honoré, and S.~Chatterjee,
\newblock ``{DANSE: Data-Driven Non-Linear State Estimation of Model-Free
  Process in Unsupervised Learning Setup},''
\newblock {\em IEEE Transactions on Signal Processing}, vol. 72, pp.
  1824--1838, 2024.

\bibitem{choPropertiesNeuralMachine2014}
K.~Cho, B.~van Merri{\"e}nboer, D.~Bahdanau, and Y.~Bengio,
\newblock ``On the properties of neural machine translation: Encoder--decoder
  approaches,''
\newblock in {\em Proc. of $8^{th}$ workshop, SSST-8}, 2014, pp. 103--111.

\bibitem{hochreiter1997long}
S.~Hochreiter and J.~Schmidhuber,
\newblock ``Long short-term memory,''
\newblock {\em Neural computation}, vol. 9, no. 8, pp. 1735--1780, 1997.

\bibitem{julier2004unscented}
S.J. Julier and J.K. Uhlmann,
\newblock ``Unscented filtering and nonlinear estimation,''
\newblock {\em Proceedings of the IEEE}, vol. 92, no. 3, pp. 401--422, 2004.

\bibitem{wan2000unscented}
E.A. Wan and R.~Van Der~Merwe,
\newblock ``The unscented {K}alman filter for nonlinear estimation,''
\newblock in {\em Proceedings of the IEEE 2000 Adaptive Systems for Signal
  Processing, Communications, and Control Symposium (Cat. No. 00EX373)}. IEEE,
  2000, pp. 153--158.

\bibitem{arasaratnam2009cubature}
I.~Arasaratnam and S.~Haykin,
\newblock ``{Cubature Kalman filters},''
\newblock {\em IEEE Transactions on Automatic Control}, vol. 54, no. 6, pp.
  1254--1269, 2009.

\bibitem{gordon1993novel}
N.~J. Gordon, D.~J. Salmond, and A.~FM. Smith,
\newblock ``Novel approach to nonlinear/non-gaussian bayesian state
  estimation,''
\newblock in {\em IEE proceedings F (radar and signal processing)}. IET, 1993,
  vol. 140, pp. 107--113.

\bibitem{arulampalam2002tutorial}
M.~S. Arulampalam, S.~Maskell, N.~Gordon, and T.~Clapp,
\newblock ``A tutorial on particle filters for online nonlinear/non-gaussian
  bayesian tracking,''
\newblock {\em IEEE Transactions on Signal Processing}, vol. 50, no. 2, pp.
  174--188, 2002.

\bibitem{sarkka2023bayesian}
S.~S{\"a}rkk{\"a} and L.~Svensson,
\newblock {\em Bayesian filtering and smoothing}, vol.~17,
\newblock Cambridge university press, 2023.

\bibitem{doucet2001sequential}
A.~Doucet, N.~De~Freitas, N.~J. Gordon, et~al.,
\newblock {\em Sequential Monte Carlo methods in practice}, vol.~1,
\newblock Springer, 2001.

\bibitem{roth2017ensemble}
Michael Roth, Gustaf Hendeby, Carsten Fritsche, and Fredrik Gustafsson,
\newblock ``The ensemble kalman filter: a signal processing perspective,''
\newblock {\em EURASIP Journal on Advances in Signal Processing}, vol. 2017,
  no. 1, pp. 56, 2017.

\bibitem{ruhe2021self}
D.~Ruhe and P.~Forr{\'e},
\newblock ``Self-supervised inference in state-space models,''
\newblock in {\em ICLR}, 2021.

\bibitem{van2025deep}
W.~L. van Nierop, N.~Shlezinger, and R.~J. van Sloun,
\newblock ``Deep variational sequential monte carlo for high-dimensional
  observations,''
\newblock in {\em ICASSP 2025-2025 IEEE International Conference on Acoustics,
  Speech and Signal Processing (ICASSP)}. IEEE, 2025, pp. 1--5.

\bibitem{corenflos2021differentiable}
A.~Corenflos, J.~Thornton, G.~Deligiannidis, and A.~Doucet,
\newblock ``Differentiable particle filtering via entropy-regularized optimal
  transport,''
\newblock in {\em International Conference on Machine Learning}. PMLR, 2021,
  pp. 2100--2111.

\bibitem{chen2023dpfsurvey}
X.~Chen and Y.~Li,
\newblock ``An overview of differentiable particle filters for data-adaptive
  sequential bayesian inference,'' 2023.

\bibitem{jonschkowski2018differentiable}
R.~Jonschkowski, D.~Rastogi, and O.~Brock,
\newblock ``Differentiable particle filters: End-to-end learning with
  algorithmic priors,''
\newblock {\em Robotics: Science and Systems XIV}, 2018.

\bibitem{girin2021dynamical}
L.~Girin, S.~Leglaive, X.~Bie, J.~Diard, T.~Hueber, and X.~Alameda-Pineda,
\newblock ``Dynamical variational autoencoders: A comprehensive review,''
\newblock {\em Foundations and Trends in Machine Learning}, vol. 15, no. 1-2,
  pp. 1--175, 2021.

\bibitem{krishnan2015deep}
R.G. Krishnan, U.~Shalit, and D.~Sontag,
\newblock ``Deep {K}alman filters,''
\newblock {\em arXiv preprint arXiv:1511.05121}, 2015.

\bibitem{krishnan2017structured}
R.~Krishnan, U.~Shalit, and D.~Sontag,
\newblock ``Structured inference networks for nonlinear state space models,''
\newblock in {\em Proceedings of the AAAI Conference on Artificial
  Intelligence}, 2017, vol.~31.

\bibitem{fraccaro2017disentangled}
M.~Fraccaro, S.~Kamronn, U.~Paquet, and O.~Winther,
\newblock ``A disentangled recognition and nonlinear dynamics model for
  unsupervised learning,''
\newblock {\em {Advances in NeurIPS}}, vol. 30, 2017.

\bibitem{ko2009gp}
J.~Ko and D.~Fox,
\newblock ``{GP-BayesFilters: Bayesian filtering using Gaussian process
  prediction and observation models},''
\newblock {\em Autonomous Robots}, vol. 27, pp. 75--90, 2009.

\bibitem{frigola2013bayesian}
R.~Frigola, F.~Lindsten, T.~B. Sch{\"o}n, and C.~E. Rasmussen,
\newblock ``Bayesian inference and learning in gaussian process state-space
  models with particle mcmc,''
\newblock {\em Advances in neural information processing systems}, vol. 26,
  2013.

\bibitem{frigola2014variational}
R.~Frigola, Y.~Chen, and C.~E. Rasmussen,
\newblock ``Variational gaussian process state-space models,''
\newblock {\em Advances in neural information processing systems}, vol. 27,
  2014.

\bibitem{svensson2016computationally}
A.~Svensson, A.~Solin, S.~S{\"a}rkk{\"a}, and T.~B. Sch{\"o}n,
\newblock ``Computationally efficient bayesian learning of gaussian process
  state space models,''
\newblock in {\em Artificial Intelligence and Statistics}. PMLR, 2016, pp.
  213--221.

\bibitem{lindsten2012ancestor}
F.~Lindsten, T.~Sch{\"o}n, and M.~Jordan,
\newblock ``Ancestor sampling for particle gibbs,''
\newblock {\em Advances in Neural Information Processing Systems}, vol. 25,
  2012.

\bibitem{lindsten2014particle}
F.~Lindsten, M.~I. Jordan, and T.~B. Sch{\"o}n,
\newblock ``Particle gibbs with ancestor sampling,''
\newblock {\em The Journal of Machine Learning Research}, vol. 15, no. 1, pp.
  2145--2184, 2014.

\bibitem{revach2022unsupervised}
G.~Revach, N.~Shlezinger, T.~Locher, X.~Ni, R.J.G. van Sloun, and Y.C. Eldar,
\newblock ``Unsupervised learned {K}alman filtering,''
\newblock in {\em 2022 30th European Signal Processing Conference (EUSIPCO)}.
  IEEE, 2022, pp. 1571--1575.

\bibitem{revach2022kalmannet}
G.~Revach, N.~Shlezinger, X.~Ni, A.~L. Escoriza, R.J.G. Van~Sloun, and Y.C.
  Eldar,
\newblock ``{KalmanNet: Neural network aided Kalman filtering for partially
  known dynamics},''
\newblock {\em IEEE Transactions on Signal Processing}, vol. 70, pp.
  1532--1547, 2022.

\bibitem{ni2024adaptive}
X.~Ni, G.~Revach, and N.~Shlezinger,
\newblock ``Adaptive kalmannet: Data-driven kalman filter with fast
  adaptation,''
\newblock in {\em ICASSP 2024 - 2024 IEEE International Conference on
  Acoustics, Speech and Signal Processing (ICASSP)}, 2024, pp. 5970--5974.

\bibitem{ghosh2025pdanse}
A.~Ghosh, Y.~C. Eldar, and S.~Chatterjee,
\newblock ``Particle-based data-driven nonlinear state estimation of model-free
  process from nonlinear measurements,''
\newblock in {\em ICASSP 2025 - 2025 IEEE International Conference on
  Acoustics, Speech and Signal Processing (ICASSP)}, 2025, pp. 1--5.

\bibitem{lorenz1963deterministic}
E.N. Lorenz,
\newblock ``Deterministic nonperiodic flow,''
\newblock {\em Journal of atmospheric sciences}, vol. 20, no. 2, pp. 130--141,
  1963.

\bibitem{ghosh2024data}
A~Ghosh, Y.~C. Eldar, and S.~Chatterjee,
\newblock ``{Semi-Supervised Model-Free Bayesian State Estimation from
  Compressed Measurements},''
\newblock {\em arXiv preprint arXiv:2407.07368}, 2025.

\bibitem{bishop2006pattern}
C.~M. Bishop and N.~M. Nasrabadi,
\newblock {\em Pattern recognition and machine learning}, vol.~4,
\newblock Springer, 2006.

\bibitem{kingma2013auto}
D.~P. Kingma and M.~Welling,
\newblock ``{Auto-Encoding Variational Bayes},''
\newblock in {\em 2nd International Conference on Learning Representations,
  {ICLR}}, 2014.

\bibitem{blanchard2021accurately}
P.~Blanchard, D.~J. Higham, and N.~J. Higham,
\newblock ``Accurately computing the log-sum-exp and softmax functions,''
\newblock {\em IMA Journal of Numerical Analysis}, vol. 41, no. 4, pp.
  2311--2330, 2021.

\bibitem{buchnik2023latent}
I.~Buchnik, G.~Revach, D.~Steger, R.~J. Van~Sloun, T.~Routtenberg, and
  N.~Shlezinger,
\newblock ``Latent-kalmannet: Learned kalman filtering for tracking from
  high-dimensional signals,''
\newblock {\em IEEE Transactions on Signal Processing}, vol. 72, pp. 352--367,
  2023.

\bibitem{li2001survey}
X.~R. Li and V.~P. Jilkov,
\newblock ``Survey of maneuvering target tracking: Iii. measurement models,''
\newblock in {\em Signal and Data Processing of Small Targets 2001}. SPIE,
  2001, vol. 4473, pp. 423--446.

\bibitem{garcia2019combining}
V.~Garcia~Satorras, Z.~Akata, and M.~Welling,
\newblock ``Combining generative and discriminative models for hybrid
  inference,''
\newblock {\em {Advances in NeurIPS}}, vol. 32, 2019.

\bibitem{paszke2019pytorch}
A.~Paszke et~al.,
\newblock ``{PyTorch: An imperative style, high-performance deep learning
  library},''
\newblock {\em {Advances in NeurIPS}}, vol. 32, 2019.

\bibitem{kingma2014adam}
D.P. Kingma and J.~Ba,
\newblock ``Adam: A method for stochastic optimization,''
\newblock in {\em 3rd International Conference on Learning Representations
  (ICLR)}, 2015.

\bibitem{lin2025efficient}
Zhidi Lin, Ying Li, Feng Yin, Juan Maro{\~n}as, and Alexandre~H Thi{\'e}ry,
\newblock ``Efficient transformed gaussian process state-space models for
  non-stationary high-dimensional dynamical systems,''
\newblock {\em IEEE Transactions on Signal Processing}, vol. 73, pp.
  5229--5243, 2025.

\bibitem{chen2022autodifferentiable}
Yuming Chen, Daniel Sanz-Alonso, and Rebecca Willett,
\newblock ``Autodifferentiable ensemble kalman filters,''
\newblock {\em SIAM Journal on Mathematics of Data Science}, vol. 4, no. 2, pp.
  801--833, 2022.

\end{thebibliography}
\end{document}